\documentclass[comment, prd,onecolumn,nofootinbib, preprintnumbers, superscriptaddress]{revtex4}

\usepackage{amsmath,amssymb,bm,slashed,braket}
\usepackage{graphicx}
\usepackage{epstopdf}
\usepackage{float}
\usepackage{xcolor}
\usepackage{multirow}
\usepackage{dsfont}
\usepackage[colorlinks=true,
            linkcolor=blue,
            urlcolor=blue,
            citecolor=green,          
            bookmarks=true,
            bookmarksnumbered=true,
            breaklinks=true,
            pdfpagemode=Fullscreen,
            pdfstartview=FitBH]{hyperref}
\usepackage[normalem]{ulem}
\allowdisplaybreaks[4]
\usepackage[capitalise]{cleveref}

\newcommand{\avec}{{\bf a}}
\newcommand{\Evec}{{\bf E}}
\newcommand{\kvec}{{\bf k}}
\newcommand{\pvec}{{\bf p}}
\newcommand{\Pvec}{{\bf P}}

\newcommand{\Rvec}{{\bf R}}
\newcommand{\xvec}{{\bf x}}
\newcommand{\yvec}{{\bf y}}
\newcommand{\bfcalR}{{\bf\cal R}}

\newcommand{\lbar}{\bar\lambda}

\newcommand{\sigmabar}{\bar\sigma}
\newcommand{\ubar}{\bar{u}}

\newcommand{\B}{\textrm{B}}
\newcommand{\BG}{\textrm{BG}}
\newcommand{\pw}{\textrm{pw}}

\newcommand{\rmL}{\textrm{L}}
\newcommand{\rmC}{\textrm{C}}
\newcommand{\rmP}{\textrm{P}}

\newcommand{\GeV}{\rm GeV}

\newcommand{\eV}{\rm eV}

\newcommand{\calA}{{\cal A}}

\newcommand{\calR}{{\cal R}}

\begin{document}

\title{All-vortex nonlinear Compton scattering in a polarized laser field}

\author{Yi Liao}
\email{liaoy@m.scnu.edu.cn}
\affiliation{School of Physics, Nankai University, Tianjin 300071, China}
\affiliation{State Key Laboratory of Nuclear Physics and
Technology, Institute of Quantum Matter, South China Normal
University, Guangzhou 510006, China}
\affiliation{Guangdong Basic Research Center of Excellence for
Structure and Fundamental Interactions of Matter, Guangdong
Provincial Key Laboratory of Nuclear Science, Guangzhou
510006, China}
\author{Quan-Yu Wang}
\email{1120200055@mail.nankai.edu.cn}
\affiliation{School of Physics, Nankai University, Tianjin 300071, China}
\author{Yuanbin Wu}
\email{yuanbin@nankai.edu.cn}
\affiliation{School of Physics, Nankai University, Tianjin 300071, China}

\begin{abstract}

The process of all-vortex nonlinear Compton scattering in an intense and polarized laser field, in which the initial and final electrons and the emitted $\gamma$ photon are all in vortex states, is studied theoretically. We develop a formalism for the process, which allows us to study the exchanges of the orbital angular momentum (OAM) and spin angular momentum among the electron, $\gamma$ photon, and laser. A wave packet of the Bessel-Gaussian type is adopted to describe the initial vortex electron for the purpose of normalization. Both circularly and linearly polarized lasers are examined. Substantial numerical calculations are performed to reveal the physics of the exchanges of the OAM and spin angular momentum. The strong impact of the laser intensity and the opening angle of the initial vortex electron is demonstrated. Our results also suggest possible scenarios for the separation of the emitted $\gamma$ photons with different OAMs, as well as a possible way which could help to distinguish the OAM and spin of the vortex particle by the multi-peak structure in the spectrum of the emitted photon.

\end{abstract}

\maketitle

\section{Introduction}

A vortex state of a particle or a twisted particle is a quantum state described by the non-plane-wave wave function with a helical phase, and carrying an intrinsic orbital angular momentum (OAM) with respect to the propagation direction \cite{Bliokh:2015doa,Knyazev:2018,Bliokh:2017uvr,Yuan:2017,Ivanov:2022jzh,Yao:11,shen:2019,Forbes:2024}. Starting from the demonstration of the optical vortex beams in the work of Allen \textit{et al.} in 1992 \cite{Allen:1992zz}, it has been shown that vortex states of photons \cite{BEIJERSBERGEN:1994,Oemrawsingh:2004,Ruffato:2014,Longman:2020,Heckenberg:1992,Peele:2002,Terhalle:2011,Gariepy:2014,Hemsing:2013,Bahrdt:2013eoa,Gauthier:2016,Lee:2019} and massive particles such as electrons \cite{Uchida:2010hbm,Verbeeck:2010ezk,McMorran:2011bql,Mafakheri:2017,Tavabi:2022,Vanacore:2019}, neutrons \cite{Clark:2015rcq,Sarenac:2019,Sarenac:2022}, and atoms \cite{luski:2021} can be produced using a number of techniques \cite{Bliokh:2017uvr,Yao:11,shen:2019,Yuan:2017,Ivanov:2022jzh,Forbes:2024}, including spiral phase plates, holographic gratings, magnetic monopole fields, and chiral plasmonic near fields. The new degree of freedom represented by the intrinsic OAM of the vortex particles, together with the development of the techniques of the production and manipulation of the vortex states, offers new opportunities for novel effects and opens up numerous applications \cite{Yao:11,Bliokh:2017uvr,shen:2019,Yuan:2017,Ivanov:2022jzh,Forbes:2024}, including quantum communications \cite{Yao:11,mair:2001,Wang:2015vrl,shen:2019,Forbes:2024} and optical manipulation \cite{Yao:11,He:1995,Simpson:1997,Grier:2003,shen:2019,Forbes:2024}. This also offers a new and powerful method \cite{Bliokh:2017uvr,Yuan:2017,Ivanov:2022jzh,Larocque:2018,kaminer:2015,madan:2020,Picón:2010,Afanasev:2016,Serbo:2015kia,Karlovets:2016uhb,Karlovets:2015nva,Maiorova:2018inm,Mandal:2020ycl,Groshev:2019uqa,Silenko:2019dfx,Hayrapetyan:2014faa,Bandyopadhyay:2015eri,Aleksandrov:2022fmp,Wu:2021trm,Lu:2023wrf,Sarenac:2024tnj,Lu:2024gha} to study and manipulate the structure of neutrons, protons, and ions, and processes of atomic, nuclear, and particle systems.

So far, the great success of vortex photons and electrons is limited to low energies \cite{Ivanov:2022jzh}. This is mainly due to the experimental demonstration and manipulation of vortex photons from visible light to x-ray (on the level of $10$ keV) \cite{shen:2019,Forbes:2024,Allen:1992zz,Longman:2020,Peele:2002,Terhalle:2011,Gariepy:2014,Hemsing:2013,Bahrdt:2013eoa,Gauthier:2016,Lee:2019} and vortex electrons with a kinetic energy up to $300$ keV \cite{Ivanov:2022jzh,Uchida:2010hbm,Verbeeck:2010ezk,McMorran:2011bql}. Inspired by this and anticipating future experimental progresses, high-energy vortex photons and electrons have been envisaged to open up new avenues in high-energy and nuclear physics \cite{Ivanov:2022jzh}. High-energy vortex photons and electrons have been shown theoretically to be a novel method to study nuclear and hadronic physics \cite{Ivanov:2022jzh,Lu:2023wrf,Lu:2024gha,Ivanov:2019vxe,Xu:2024jlt,Afanasev:2021fda}, and probe the quantum electrodynamics (QED) effects and fundamental aspects of vortex states \cite{Ivanov:2022jzh,Bu:2021ebc,Lei:2021eqe,Sherwin:2017vsf,Maruyama:2017ptl,Bliokh:2017sdz,Silenko:2019okz,Silenko:2018eed,vanKruining:2017anw,Barnett:2017wrr,Bialynicki-Birula:2016unl}. However, the production as well as the detection of high-energy vortex particles so far still remains a challenge \cite{Ivanov:2022jzh,Bu:2024,Li:2024mzd}, as traditional techniques which work for low energies become impractical at high energies due to problems such as damage thresholds and fabrication complexities. The present proposals for the production of high-energy vortex photons are mainly based on scattering processes, including Compton scattering \cite{Katoh:2016yqc,Petrillo:2016,Jentschura:2010ap,Jentschura:2011ih,Katoh:2016aww}, nonlinear Compton scattering in which intense lasers are involved \cite{Taira:2017,Chen:2018tkb,Ababekri:2022mob,Guo:2023uyu,Bogdanov:2019ocq,Jiang:2024fit}, and laser-plasma interactions \cite{Wang:2020,liu:2016,hu:2021}. For the production of high-energy vortex electrons, similar methods based on the nonlinear Compton scattering process or laser-plasma interactions have been proposed \cite{Bu:2024,ju:2018}, as well as the method of accelerating vortex electrons by the axially symmetric fields of electric and magnetic lenses which could preserve the angular momentum and the shape of a wave packet \cite{Karlovets:2020tlg} and the method via generalized measurements or quantum entanglement \cite{Karlovets:2022evc,Karlovets:2022mhb}.

The process of nonlinear Compton scattering has been considered to be one of promising ways for the production and investigation of high-energy vortex photons and electrons \cite{Taira:2017,Chen:2018tkb,Ababekri:2022mob,Guo:2023uyu,Bogdanov:2019ocq,Bu:2024,Jiang:2024fit}. However, a full quantum treatment of the process with vortex states has thus far rather limited~\cite{Ababekri:2022mob,Guo:2023uyu,Bu:2024}. Here, we study the all-vortex nonlinear Compton scattering in a polarized laser field, in which the initial and final electrons and the emitted $\gamma$ photon are all in a vortex state. We develop in the Furry picture of QED a theoretical formalism for the process, which allows us to study the exchanges of the OAM and spin angular momentum among the electron, $\gamma$ photon, and laser. We adopt a wave packet of the Bessel-Gaussian (BG) type to describe the initial vortex electron for the purpose of normalization, and consider both circularly and linearly polarized lasers. We perform numerical calculations for a relativistic electron, to study the effects of the OAM and spin of the involved particles and the opening angle of the initial vortex electron, as well as the laser intensity. Possible scenarios for the separation of the emitted $\gamma$ photons with different OAMs are analyzed, as well as the multi-peak structure in the emitted photon spectrum which could help to identify the OAM and spin of the vortex particle.

This article is organized as follows. In \cref{sec:analytical}, we present the theoretical formalism for the process of all-vortex nonlinear Compton scattering in which the initial electron is normalized by a BG state. We proceed with differential rates as analytically as we could. Our numerical results are demonstrated in \cref{sec:numerical} by various distributions of the emitted vortex $\gamma$ photon. Possible scenarios for the separation of emitted $\gamma$ photons with different OAMs, the features of their spectra, and the exchanges of the OAM and spin angular momentum among the electrons, $\gamma$ photon, and laser are also discussed. We finally conclude in \cref{sec:conclusion}. The natural units ($\hbar=c=1$) are used throughout the whole article, and $\alpha=e^2/(4\pi)\sim 1/137$ is the fine structure constant with $e$ being the electron charge.

\section{Analytical calculation}
\label{sec:analytical}

\subsection{Vortex states and their normalization}
\label{sec:states}

We begin with the definitions and conventions for the process under consideration: a vortex electron collides head-on with a monochromatic, plane-wave, strong laser to radiate a vortex photon while continuing to stay in a vortex state. We assume the laser propagates in the $(-z)$ axis with the four-momentum $\Omega^\mu=\omega(1,0,0,-1)$, where $\omega$ is its angular frequency. The laser is either linearly polarized, which we choose without loss of generality to be in the $x$ axis, 
\begin{eqnarray}
a^\mu(\tau)=M\xi\epsilon_1^\mu\cos\tau,
\label{eq:laser_linear}
\end{eqnarray}
where $\tau=\Omega\cdot x$ and $\epsilon_1^\mu=(0,1,0,0)$ (and $\epsilon_2^\mu=(0,0,1,0)$ for the following use), or circularly polarized which we choose to be left-handed, 
\begin{eqnarray}
a^\mu(\tau)=\frac{1}{\sqrt{2}}M\xi\left(\epsilon_1^\mu\cos\tau+\epsilon_2^\mu\sin\tau\right).
\label{eq:laser_circular}
\end{eqnarray}
Here $M$ is the electron mass, and $\xi$ is a dimensionless parameter so that the electric field strength has the amplitude $E=E_\textrm{cr}\xi\omega/M=M\xi\omega/|e|$ with $E_\textrm{cr}=M^2/|e|$ being the Schwinger critical field strength. We have chosen for both polarizations the same time-average of the electric field squared, $\overline{\Evec^2}=(M\xi\omega/e)^2/2$, or $|\overline{a^2}|=(M\xi)^2/2$. 

\subsubsection{Bessel vortex state}
\label{sec:bessel}

In this work we consider first the case in which all of the initial and final electrons and the final photon are in an ideal Bessel vortex state that is the simplest one that exhibits essential features of a vortex. Since the Bessel vortex is not square-integrable, we will incorporate a Gaussian smearing to the initial electron vortex, i.e., the so-called Bessel-Gaussian vortex, so that the scattering rate can be well defined. In the future work, we will investigate a scenario that would be closer to the experimental situation: the initial electron in a Laguerre-Gauss state collides against a counter-propagating laser pulse that has a finite transverse distance (impact parameter) to the electron. 

The kernel of a Bessel vortex is, choosing the $z$ axis as the longitudinal direction \cite{Ivanov:2011kk}, 
\begin{eqnarray}
a_{\kappa\ell}(\pvec_\perp)=i^{-\ell}e^{i\ell\phi_p}\sqrt{\frac{2\pi}{\kappa}}\delta(p_\perp-\kappa),~~~p_\perp=|\pvec_\perp|,
\end{eqnarray}
where $\ell$ is an integer that will become in the coordinate representation the projection of OAM in the $z$ axis and offers the azimuthal phase essential to a vortex state, and $\kappa$ is a specified magnitude of the transverse momentum. We denote the components of a vector in the transverse plane by a subscript $\perp$. Then a Bessel vortex electron that propagates in the $z$ axis has the wavefunction in free space: 
\begin{eqnarray}
\psi(x)
=e^{i(p_zz-Et)}\int\frac{d^2\pvec_\perp}{(2\pi)^2}a_{\kappa\ell}(\pvec_\perp)e^{i\pvec_\perp\cdot\xvec_\perp}u(p,\sigma).
\end{eqnarray}
Here $u(p,\sigma)$ is the usual bispinor that can be built by 
$u(p,\sigma)=(E+M)^{-1/2}(M+\slashed{p})u({\bf 0},\sigma)$, where $u({\bf 0},\sigma)$ (with $\sigma=\pm$) is the normalized bispinor for an electron in its rest frame with spin projection $\sigma/2$ in the $z$ direction. In the Dirac or standard representation \cite{Laudau:1982}, 
\begin{subequations}
\begin{eqnarray}
&&u(p,\sigma)
=\left(\begin{array}{l}
\sqrt{E+M}w_\sigma \\
\sqrt{E-M}\hat\pvec\cdot\vec\sigma w_\sigma
\end{array}\right),
\\
&&w_+=\left(
\begin{array}{r}
1 \\ 0
\end{array}\right),~~~
w_-=\left(
\begin{array}{r}
0 \\ 1
\end{array}\right),
\end{eqnarray}
\end{subequations}
with $\sigma^3w_\sigma=\sigma w_\sigma$. Note that $u$ is normalized by $\ubar u=2M$, and we will handle the normalization of the vortex state soon. 

The presence of a plane-wave laser background specifies a direction in spacetime, thus not all Descartes components of momentum are conserved between the initial and final states. Following the usual practice, we work with the light-front coordinates to better preserve energy-momentum conservation\cite{Brodsky:1997de}. As we have assumed the laser propagates in the $(-z)$ axis, we denote for any four-vector $x^\mu=(x^0,\xvec_\perp,x^3)$, 
\begin{eqnarray}
x_\pm=x^0\pm x^3,
\end{eqnarray}
where from now on we will use the superscripts and subscripts for the $\pm$ components arbitrarily, so that the scalar product, integration measure, and $\delta$ functions between the Descartes and light-front coordinates are related by 
\begin{subequations}
\begin{eqnarray}
&&x\cdot y=x^0y^0-\xvec\cdot\yvec
=\frac{1}{2}x_+y_-+\frac{1}{2}x_-y_+-\xvec_\perp\cdot\yvec_\perp,
\\
&&dx^0d^3\xvec=\frac{1}{2}dx_+dx_-d^2\xvec_\perp,
\\
&&\delta^4(x-y)=\delta(x^0-y^0)\delta^3(\xvec-\yvec)
=2\delta(x_+-y_+)\delta(x_--y_-)\delta^2(\xvec_\perp-\yvec_\perp). 
\end{eqnarray}
\end{subequations}
In particular the on-shell condition for a particle of momentum $p$ and mass $M$ is, $p^2=p_+p_--\pvec_\perp^2=M^2$. Employing the light-front coordinates and noting that conservation of the plus component of momentum is not affected by the presence of the laser due to $\Omega_+=0$, 
the Bessel-vortex state for the radiated photon can be designated by $k_+$ among other quantities: 
\begin{eqnarray}
A^\mu_{k_+\kappa_\gamma\ell_\gamma\lambda}(x)
&=&e^{-i(k_+x_- +k_-x_+)/2}\int\frac{d^2\kvec_\perp}{(2\pi)^2}
a_{\kappa_\gamma\ell_\gamma}(\kvec)e^{i\kvec_\perp\cdot\xvec_\perp}\Lambda^\mu(k,\lambda).
\label{eq:photon_bessel}
\end{eqnarray}
Here $\Lambda^\mu(k,\lambda)$ is the polarization vector for a plane-wave photon of momentum $k$ and helicity $\lambda$, which cannot be taken outside of the integral. It is convenient to parameterize it with the help of $\epsilon_\lambda$ and $\Omega$ \cite{Seipt:2020diz,King:2020btz}: 
\begin{eqnarray}
\Lambda(k,\lambda)=\epsilon_\lambda^*-\frac{k\cdot\epsilon^*_\lambda}{k\cdot\Omega}\Omega,
\end{eqnarray}
where $\epsilon_\lambda=(\epsilon_1+i\lambda\epsilon_2)/\sqrt{2}$ with $\Lambda(k,\lambda)\cdot k=0$ and $\Lambda^*(k,\lambda_1)\cdot\Lambda(k,\lambda_2)=-\delta_{\lambda_1\lambda_2}$. And $\kappa_\gamma$ is the magnitude of its transverse momentum and the integer $\ell_\gamma$ denotes the projection of its angular momentum in the longitudinal direction. 

When immersed in a strong plane-wave laser, the so-called Bessel-Volkov solution to the Dirac equation takes into account the laser field exactly. The wavefunctions for the initial and final electrons are denoted respectively by \cite{Hayrapetyan:2014faa,Karlovets:2012eu}
\begin{subequations}
\begin{eqnarray}
\psi_{p_+\kappa\ell\sigma}^{(a)}(x)
&=&e^{-i(p_+x_- +p_-x_+)/2}\int\frac{d^2\pvec_\perp}{(2\pi)^2}a_{\kappa\ell}(\pvec_\perp)e^{i\pvec_\perp\cdot\xvec_\perp}
e^{i\Phi_p(\tau)}
\left[1+\frac{\slashed{\Omega}\slashed{a}}{2p\cdot\Omega}\right]u(p,\sigma),
\label{eq:initial_e}
\\
\psi_{p'_+\kappa'\ell'\sigma'}^{(a)}(x)
&=&e^{-i(p'_+x_- +p'_-x_+)/2}\int\frac{d^2\pvec'_\perp}{(2\pi)^2}a_{\kappa'\ell'}(\pvec'_\perp)e^{i\pvec'_\perp\cdot\xvec_\perp}
e^{i\Phi_{p'}(\tau)}\left[1+\frac{\slashed{\Omega}\slashed{a}}{2p'\cdot\Omega}\right]u(p',\sigma'),
\end{eqnarray}
\end{subequations}
where the superscript $(a)$ indicates the presence of the laser background $a^\mu(\tau)$ which introduces an additional phase, 
\begin{eqnarray}
&&\Phi_p(\tau)=\int^\tau d\tau'\left[-\frac{p\cdot a(\tau')}{p\cdot\Omega}+\frac{a^2(\tau')}{2p\cdot\Omega}\right].
\end{eqnarray}
Here $\kappa$ ($\kappa'$) is the magnitude of the transverse momentum of the initial (final) electron and the integer $\ell$ ($\ell'$) denotes the longitudinal projection of its angular momentum. 

Now we consider normalization and related issues. The normalization of a Bessel vortex state in free space has been well discussed in the literature \cite{Ivanov:2011kk,Jentschura:2011ih,Ivanov:2011tm}. In the presence of a laser background the normalization of a Bessel vortex state for a neutral particle is obtained by replacing Descartes components by light-front components; for instance, for the radiated Bessel photon, the normalization factor $N_k^\B$ to be multiplied to Eq.~\eqref{eq:photon_bessel}, the number density of states in the interval $dk_\perp dk_+$, and the integration measure in the final-state phase space are respectively: 
\begin{subequations}
\begin{eqnarray}
&&N_k^\B=\frac{1}{\sqrt{2k_+}}\sqrt{\frac{2\pi}{RL_-}},
\label{eq:norm_free}
\\
&&dn_k=\frac{Rdk_\perp}{\pi}\frac{L_-dk_+}{4\pi},
\\
&&(N_k^\B)^2dn_k=\frac{dk_\perp dk_+}{2\pi 2k_+}.
\end{eqnarray}
\end{subequations}
Here $L_-$ is a large length of the coordinate component that is conjugate to the conserved plus component of momentum, and $R$ is a large enough radius in the transverse plane beyond which the wavefunction almost vanishes. Since the vortex state of a charged particle is modified by a strong laser field, we have to recalculate its normalization. For the Bessel-Volkov solution of the electron in Eq.~\eqref{eq:initial_e}, we calculate the plus component of its current in \cref{sec:appendix_A}, and it turns out that it yields the same normalization constant as in free space. 

\subsubsection{Bessel-Gaussian vortex state}
\label{sec:gaussian}

Finally we discuss the Bessel-Gaussian vortex for the initial electron which is square-integrable in the transverse plane, and we will employ it for numerical analysis. It is an integral of the wavefunction in Eq.~\eqref{eq:initial_e} over the magnitude $\kappa$ of the transverse momentum weighted by the Gaussian function $f(\kappa;\kappa_0,\sigma_\perp)$ with the central value $\kappa_0$ and deviation $\sigma_\perp$ \cite{Ivanov:2011bv}:
\begin{eqnarray}
\psi_{p_+\kappa_0\sigma_\perp\ell\sigma}^{{(a)}\BG}(x)
=\int_0^\infty d\kappa~\psi_{p_+\kappa\ell\sigma}^{(a)}(x)f(\kappa;\kappa_0,\sigma_\perp),
\label{eq:initial_eBG}
\end{eqnarray}
where 
\begin{eqnarray}
f(\kappa;\kappa_0,\sigma_\perp)
=N(\kappa_0,\sigma_\perp)\exp\bigg[-\frac{(\kappa-\kappa_0)^2}{2\sigma_\perp^2}\bigg],
\label{eq:gaussian}
\end{eqnarray}
whose square normalization to unity yields 
\begin{eqnarray}
N^{-2}(\kappa_0,\sigma_\perp)=\frac{1}{2}\sqrt{\pi}\sigma_\perp\left[1+\textrm{erf}\left(\frac{\kappa_0}{\sigma_\perp}\right)\right].
\end{eqnarray}
Since we require that the Bessel-Gaussian vortex be on-shell and $p_+$ be a good quantum number in the presence of a laser, smearing in $\kappa$ implies variation in $p_-$ according to $p_+p_--\kappa^2=M^2$. 

The issue of normalization has to be reconsidered for a Bessel-Volkov solution whose magnitude of transverse momentum is Gaussian smeared, i.e., Eq.~\eqref{eq:initial_eBG}, where the Bessel-Volkov solution $\psi_{p_+\kappa\ell\sigma}^{(a)}$ is expressed in Eq.~\eqref{eq:psi_lightfront} (without the normalization factor $N_p^\B$ of course). When forming the plus component of the current, $j_+=\bar\psi_{p_+\kappa_0\sigma_\perp\ell\sigma}^{(a)\BG}\gamma_+\psi_{p_+\kappa_0\sigma_\perp\ell\sigma}^{(a)\BG}$, we notice several points that are different from the pure Bessel-Volkov case. We denote by a tilde all momentum-dependent quantities in $\bar\psi_{p_+\kappa_0\sigma_\perp\ell\sigma}^{(a\BG)}$; for instance, the Gaussian integration variable in $\bar\psi_{p_+\kappa_0\sigma_\perp\ell\sigma}^{(a\BG)}$ is $\tilde\kappa$. With a fixed $p_+$, $p\cdot\Omega=p_+\Omega_-/2$ is also fixed. This results in simplifications in $j_+$. For instance, the phase $\phi_a^p$ keeps intact from smearing and the variable $\bfcalR_\perp$ is the same in both $\bar\psi_{p_+\kappa_0\sigma_\perp\ell\sigma}^{(a)\BG}$ and $\psi_{p_+\kappa_0\sigma_\perp\ell\sigma}^{(a)\BG}$; for relevant definitions, see \cref{sec:appendix_A}. We obtain 
\begin{eqnarray}
j_+(x)&=&\int d\kappa\int d\tilde\kappa~
f\tilde f e^{i(\tilde p_--p_-)x_+/2}
\sqrt{\frac{\kappa\tilde\kappa}{(2\pi)^2}}
\nonumber
\\
&&\times
\Big[(\tilde\epsilon_+ +\tilde\epsilon_-\tilde c)(\epsilon_++\epsilon_-c)
J_\ell(\kappa\calR_\perp) J_\ell(\tilde\kappa\calR_\perp)
+\tilde\epsilon_-\epsilon_- \tilde s s
J_{\ell+\sigma}(\kappa\calR_\perp) J_{\ell+\sigma}(\tilde\kappa\calR_\perp)\Big],
\end{eqnarray}
where $c=\cos\theta_p$, $\tilde s=\sin\theta_{\tilde p}$, etc. Using \cite{Ponce_de_Leon:2015}:
\begin{eqnarray}
\int_0^\infty xdx~J_\ell(\kappa x)J_\ell(\kappa'x)=\frac{1}{\kappa}\delta(\kappa-\kappa'),
\end{eqnarray}
and transforming integration over $\xvec_\perp$ to that over $\bfcalR_\perp$ as in \cref{sec:appendix_A}, all complications arising from smearing disappear: 
\begin{eqnarray}
\int\frac{1}{2}dx^-d^2\xvec_\perp~j_+(x)
=\frac{L_-}{2}\int 2\pi\calR_\perp d\calR_\perp~j_+(x)
=\frac{L_-}{2}2p^+. 
\end{eqnarray}
This is as nice as anticipated: the state is normalized in the transverse plane of space and is normalized as for a plane wave in the minus axis. In summary, for the Bessel photon and Bessel-Volkov electron states we employ the normalization factor as in Eq.~\eqref{eq:norm_free} and for the initial Bessel-Volkov electron whose magnitude of transverse momentum is smeared we employ the normalization factor:
\begin{eqnarray}
N_p^\BG=\frac{1}{\sqrt{2p_+}}\sqrt{\frac{2}{L_-}}. 
\label{eq:nor_gauss}
\end{eqnarray}

\subsection{Scattering amplitude for plane-wave electrons and photon}
\label{sec:plane_wave}

Since a vortex state is a linear composition of infinitely many plane-wave states, the $S$ matrix element for a process involving vortex states, $S_\B$, can be constructed from that involving only plane-wave states, $S_\pw$. For the simplest case of pure Bessel-vortex states and without taking into account differences in normalization for the moment, we have for the process under consideration \cite{Ivanov:2011kk}: 
\begin{eqnarray}
S_\B&=&\int\frac{d^2\pvec_\perp}{(2\pi)^2}
\int\frac{d^2\pvec'_\perp}{(2\pi)^2}
\int\frac{d^2\kvec_\perp}{(2\pi)^2}
a^\dagger_{\kappa_\gamma\ell_\gamma}(\kvec_\perp)
a^\dagger_{\kappa'\ell'}(\pvec'_\perp)
a_{\kappa\ell}(\pvec_\perp)S_\pw,
\label{eq:amplitude_relation}
\end{eqnarray}
where 
\begin{subequations}
\begin{eqnarray}
S_\pw
&=&-ie\int d^4x~e^{-iP\cdot x}e^{i[\Phi_p(\tau)-\Phi_{p'}(\tau)]}U,
\\
U&=&\Lambda^\mu(k,\lambda)\overline{u(p',\sigma')}
\left[1+\frac{\slashed{a}\slashed{\Omega}}{2p'\cdot\Omega}\right]\gamma_\mu
\left[1+\frac{\slashed{\Omega}\slashed{a}}{2p\cdot\Omega}\right]u(p,\sigma),
\end{eqnarray}
\end{subequations}
and $P=p-p'-k$. In the following we first compute the bilinear form $U$, then treat the phases and finish integration over the coordinate $x$. 

\subsubsection{Bilinear}
\label{sec:bilinear}

Since the laser quadratic term does not survive using $\slashed{\Omega}\slashed{\epsilon}_\lambda^*\slashed{\Omega}=0$ and $\slashed{\Omega}\slashed{\Omega}=0$, the bilinear becomes, 
\begin{eqnarray}
U_\lambda&=&\big[2(E+M)(E'+M)\big]^{-1/2}
\big(B_0+\sqrt{2}\avec_\perp\cdot\vec\epsilon_\lambda^*B_1\big), 
\end{eqnarray}
where for the electron spin-preserved case ($\sigma'=\sigma$), 
\begin{subequations}
\begin{eqnarray}
B_0&=&-2\delta_{\lambda\sigmabar}(E'+M)p_\sigma
-2\delta_{\lambda\sigma}(E+M)p'_{\sigmabar}
\nonumber
\\
&&+r_\lambda\Big[(p'_+ +M)(p_+ +M)
+(\pvec'_\perp\cdot\pvec_\perp+i\sigma R^3)\Big],
\\
B_1&=&+\bigg(\frac{\delta_{\lambda\sigma}}{p'_+}
+\frac{\delta_{\lambda\sigmabar}}{p_+}\bigg)(p'_+ +M)(p_+ +M)
\nonumber
\\
&&+\bigg(\frac{\delta_{\lambda\sigmabar}}{p'_+}
+\frac{\delta_{\lambda\sigma}}{p_+}\bigg)(\pvec'_\perp\cdot\pvec_\perp+i\sigma R^3),
\end{eqnarray}
\end{subequations}
and for the electron spin-flipped case ($\sigma'=-\sigma$),
\begin{subequations}
\begin{eqnarray}
B_0&=&2\sigma\delta_{\lambda\sigma}
\Big[(p_+ +M)p^{\prime 3}-(p'_+ +M)p^3\Big]
\nonumber
\\
&&+r_\lambda\sigma\Big[(p_+ +M)p'_\sigma-(p'_+ +M)p_\sigma\Big],
\\
B_1&=&
-\sigma\bigg(\frac{\delta_{\lambda\sigmabar}}{p'_+}
+\frac{\delta_{\lambda\sigma}}{p_+}\bigg)(p'_+ +M)p_\sigma
\nonumber
\\
&&+\sigma\bigg(\frac{\delta_{\lambda\sigma}}{p'_+}
+\frac{\delta_{\lambda\sigmabar}}{p_+}\bigg)(p_+ +M)p'_\sigma.
\end{eqnarray}
\end{subequations}
Both $B_0$ and $B_1$ are independent of the laser parameters which enter only in the global factor $\sqrt{2}\avec_\perp\cdot\vec\epsilon_\lambda^*$. Note that the laser polarization is determined by $\avec_\perp$ while $\epsilon_\lambda^\mu$ is an auxiliary vector that depends on the helicity $\lambda$ of the radiated photon. We have introduced the following abbreviations:
\begin{subequations}
\begin{eqnarray}
&&
\lbar=-\lambda,~~~\sigmabar=-\sigma,~~~
r_\lambda=\frac{1}{k_+}\kvec_\perp\cdot\sqrt{2}\vec\epsilon_\lambda^*,
\\
&&
\Rvec=\pvec'\times\pvec,~~~R_\sigma=R^1+i\sigma R^2,
\\
&&
p_\sigma=p^1+i\sigma p^2,~~~
p'_\sigma=p^{\prime 1}+i\sigma p^{\prime 2}.
\end{eqnarray}
\end{subequations}

\subsubsection{Phases}
\label{sec:phases}

There are two types of phases; one is from the usual plane-wave: 
\begin{eqnarray}
-P\cdot x=-\frac{1}{2}P_+x_- -\frac{1}{2}P_-x_++\Pvec_\perp\cdot\xvec_\perp,
\end{eqnarray}
and the other is introduced by the Volkov solution, which becomes after some algebra:  
\begin{eqnarray}
&&\Phi^{(a)}=\Phi_p(\tau)-\Phi_{p'}(\tau)
=\int^\tau d\tau'\left[-\frac{k\cdot a}{p'\cdot\Omega}
+\frac{k\cdot\Omega}{p'\cdot\Omega p\cdot\Omega}\left(p\cdot a-\frac{1}{2}a^2\right)
\right].
\end{eqnarray}
The laser introduces nontrivial $x$ dependence through $\tau=\Omega\cdot x=\Omega^-x^+/2=\omega(x^0+x^3)$. Thus the $x$ components other than $x^+$ can be finished to the $\delta$ functions as usual; in other words, the minus components of the particles' momenta are not conserved due to their exchange with the laser while the other three components are still conserved. To incorporate all components in a uniform manner, we follow Ref. \cite{Seipt:2020diz} to introduce a parameter integral: 
\begin{eqnarray}
\int\Omega^- du~\delta\left(u\Omega^-+P^-\right)=1.
\end{eqnarray}
Since all other components of $\Omega$ vanish, the above $\delta$ function can be combined with the other three from integration over $x^-$ and $\xvec_\perp$ into 
\begin{eqnarray}
\delta^4\left(u\Omega+P\right)
=2\delta\left(u\Omega^-+P^-\right)\delta\left(P^+\right)\delta^2\left(\Pvec_\perp\right),
\end{eqnarray}
which yields three equivalent answers for $u$:
\begin{eqnarray}
u=-\frac{P^-}{\Omega^-}=\frac{k\cdot p}{p'\cdot\Omega}=\frac{k\cdot p'}{p\cdot\Omega}.
\label{eq:u}
\end{eqnarray}
To summarize what we have achieved so far, 
\begin{eqnarray}
S_\pw
&=&-ie
\int\frac{du}{2\pi}\int d\tau~(2\pi)^4 \delta^4\left(u\Omega+P\right) 
e^{i[u\tau+\Phi^{(a)}]}U.
\end{eqnarray}

To proceed further, we have to work out the laser-dependent phase $\Phi^{(a)}$. For the circularly polarized laser in Eq.~\eqref{eq:laser_circular}, we have 
\begin{eqnarray}
\Phi^{(a)}=2\beta\tau -\gamma^\textrm{C}\sin(\tau-\tau_\rho),
\end{eqnarray}
where the following notations are introduced,
\begin{subequations}
\begin{eqnarray}
&&\beta=\frac{k_+(M\xi)^2}{8p_+p'_+\omega},~~~
\gamma^\textrm{C}=\frac{p_+M\xi}{\sqrt{2}p'_+\omega}\rho,~~~
\label{eq:beta_gammaC}
\\
&&\vec\rho=\frac{k_+}{p_+}\frac{\pvec_\perp}{p_+}-\frac{\kvec_\perp}{p_+}
=\rho(\cos\tau_\rho,\sin\tau_\rho),
\label{eq:rho}
\end{eqnarray}
\end{subequations}
and for the linearly polarized laser in Eq.~\eqref{eq:laser_linear}, we obtain 
\begin{eqnarray}
\Phi^{(a)}=2\beta\tau-\gamma^\rmL\sin\tau+\beta\sin(2\tau), 
\end{eqnarray}
where 
\begin{eqnarray}
\gamma^\rmL=\frac{M\xi}{p'\cdot\Omega}\left[
k\cdot\epsilon_1 -\frac{k\cdot\Omega}{p\cdot\Omega}p\cdot\epsilon_1\right]
=\sqrt{2}\gamma^\rmC\cos\tau_\rho.
\end{eqnarray}

\subsubsection{Plane-wave amplitude}
\label{sec:amplitude_pw}

To facilitate the $\tau$ integral, we notice that the triangular terms in $\Phi^{(a)}$ are periodic functions of $\tau$, so that their phase factors should better be expanded in integer powers of $e^{i\tau}$. By using \cite{Gradshteyn:2007} 
\begin{eqnarray}
e^{iz\sin\theta}=\sum_{n=-\infty}^\infty J_n(z)e^{in\theta}, 
\end{eqnarray}
we have for the circularly polarized laser, 
\begin{subequations}
\begin{eqnarray}
e^{-i\gamma^\rmC\sin(\tau-\tau_\rho)}
&=&\sum_{n=-\infty}^\infty J_n(\gamma^\rmC)e^{-in\tau}e^{in\tau_\rho},
\\
e^{\pm i\tau}e^{-i\gamma^\rmC\sin(\tau-\tau_\rho)}
&=&\sum_{n=-\infty}^\infty J_{n\pm 1}(\gamma^\rmC)e^{-in\tau}e^{i(n\pm 1)\tau_\rho},
\end{eqnarray}
\end{subequations}
and for the linearly polarized laser, 
\begin{subequations}
\begin{eqnarray}
e^{i[-\gamma^\rmL\sin\tau+\beta\sin(2\tau)]}
&=&\sum_{n,\nu=-\infty}^\infty e^{-in\tau}J_{2\nu+n}(\gamma^\rmL)J_\nu(\beta),
\\
e^{i[-\gamma^\rmL\sin\tau+\beta\sin(2\tau)]}\cos\tau
&=&\frac{1}{2}\sum_{n,\nu=-\infty}^\infty e^{-in\tau}\left[J_{2\nu+n-1}(\gamma^\rmL) +J_{2\nu+n+1}(\gamma^\rmL)\right]J_\nu(\beta),
\end{eqnarray}
\end{subequations}
where the $e^{\pm i\tau}$ or $\cos\tau$ factor is anticipated in the linear $a$ term of the bilinear $U$. 

The $\tau$ integral can now be finished to yield $\delta(u+2\beta-n)$ for both polarizations of the laser, which in turn is used to finish the $u$ integral in terms of $u_n=n-2\beta$. Since $u+2\beta>0$, the summation over $n$ is actually restricted to $n\ge n_0$ where $n_0$ is some positive integer. The scattering matrix in the plane-wave case becomes finally, 
\begin{eqnarray}
S_\pw^\rmP
&=&-ie\frac{(2\pi)^4 2\delta(P^+)\delta^2(\Pvec_\perp)}{\sqrt{2(E+M)(E'+M)}}
\sum_{n=n_0}^{+\infty}\delta(u_n\Omega^-+P^-)
\Big(\calA^\rmP_{0n}+|a|\calA^\rmP_{1n}\Big), 
\label{eq:amplitude_pw}
\end{eqnarray}
where the superscript $\rmP$ refers to the case of a linearly polarized laser $\rmP=\rmL$, 
\begin{subequations}
\begin{eqnarray}
&&\calA^\rmL_{0n}
=B_0\sum_{\nu=-\infty}^\infty
J_\nu(\beta)J_{2\nu+n}(\gamma^\textrm{L}),
\\
&&\calA^\rmL_{1n}=\frac{1}{2}B_1\sum_{\nu=-\infty}^\infty
J_\nu(\beta)\Big(J_{2\nu+n-1}(\gamma^\textrm{L}) +J_{2\nu+n+1}(\gamma^\textrm{L})\Big),
\end{eqnarray}
\end{subequations}
or the case of a circularly polarized laser $\rmP=\rmC$, 
\begin{subequations}
\begin{eqnarray}
&&\calA^\rmC_{0n}
=B_0J_n(\gamma^\rmC)e^{in\tau_\rho},
\\
&&
\calA^\rmC_{1n}
=\frac{1}{\sqrt{2}}B_1 J_{n-\lambda}(\gamma^\rmC)e^{i(n-\lambda)\tau_\rho}.
\end{eqnarray}
\end{subequations}

\subsection{Scattering amplitude for Bessel-vortex electrons and photon}
\label{sec:Bessel_vortex}

By using Eqs.~\eqref{eq:amplitude_relation} and \eqref{eq:amplitude_pw}, the $S$ matrix in the Bessel-vortex case becomes 
\begin{eqnarray}
&&S_\B^\rmP
=-ie\sum_{n=n_0}^\infty
\int\frac{d^2\pvec_\perp}{(2\pi)^2}
\int\frac{d^2\pvec'_\perp}{(2\pi)^2}
\int\frac{d^2\kvec_\perp}{(2\pi)^2}
(2\pi)^4\delta^4(u_n\Omega+P)
F^\rmP(\varphi,\varphi',\varphi_k),
\end{eqnarray}
where the normalization factors and $[2(E+M)(E'+M)]^{-1/2}$ on the right-hand side will be finally resumed. To emphasize dependence on the azimuthal angles $\varphi$ for $\pvec_\perp$, $\varphi'$ for $\pvec'_\perp$, and $\varphi_k$ for $\kvec_\perp$, we denote 
\begin{eqnarray}
F^\rmP(\varphi,\varphi',\varphi_k)
&=&a^\dagger_{\kappa_\gamma\ell_\gamma}(\kvec_\perp)
a^\dagger_{\kappa'\ell'}(\pvec'_\perp)
a_{\kappa\ell}(\pvec_\perp)
\Big(\calA^\rmP_{0n}+M\xi\calA^\rmP_{1n}\Big).
\end{eqnarray}
Now we work out the integrals over the three transverse momenta. 

\subsubsection{Transverse integrals excluding $\varphi$}
\label{sec:transverse}

The integral over the magnitude of each transverse momentum is trivially finished by the $\delta$ function of the Bessel vortex. The planar integrals over any two of the three azimuthal angles, say, $\varphi',~\varphi_k$, with the constraint of transverse momentum conservation are accomplished in Appendix \ref{sec:appendix_B}. The result is, 
\begin{eqnarray}
S_\B^\rmP
&=&-ie\sum_{n=n_0}^\infty
(2\pi)^4 2\delta(u_n\Omega^- +P^-)\delta(P^+)
\int\frac{p_\perp dp_\perp}{(2\pi)^2}
\int\frac{p'_\perp dp_\perp}{(2\pi)^2}
\int\frac{k_\perp dk_\perp}{(2\pi)^2}\frac{1}{2\Delta}
\int d\varphi\sum_{\eta=\pm}
F^\rmP(\varphi,\varphi'_\eta,\varphi_{k\eta}),
\end{eqnarray}
where $\eta=\pm$ denotes the two solutions to the constraint on three transverse momenta of fixed magnitudes, and 
\begin{subequations}
\begin{eqnarray}
&&\Delta=\frac{1}{4}\sqrt{-\lambda(p_\perp^2,p_\perp^{\prime 2},k_\perp^2)},
\\
&&\varphi'_\eta=\varphi -\eta\delta',~~~
\varphi_{k\eta}=\varphi +\eta\delta_k,
\\
&&\angle(\pvec_\perp,\pvec'_\perp)
=\delta'=\arccos\frac{p_\perp^2+p_\perp^{\prime 2}-k_\perp^2}{2p_\perp p'_\perp},
\label{angle between electron and electron}
\\
&&\angle(\pvec_\perp,\kvec_\perp)
=\delta_k=\arccos\frac{p_\perp^2+k_\perp^2-p_\perp^{\prime 2}}{2p_\perp k_\perp},
\label{angle between electron and photon}
\end{eqnarray}
\end{subequations}
where we used the standard triangular function $\lambda(a,b,c)=a^2+b^2+c^2-2ab-2bc-2ca$. 

\subsubsection{Integrals over $\varphi$}
\label{sec:varphi_integrals}

Inspecting the form of $F^\rmP(\varphi,\varphi'_\eta,\varphi_{k\eta})$, we see that there are three sources of $\varphi$ dependence. First, the kernel of the Bessel vortex for all three states offers the phase: 
\begin{eqnarray}
e^{i(\ell\varphi-\ell'\varphi'_\eta-\ell_\gamma\varphi_{k\eta})}
=e^{i\varphi\Delta\ell}e^{i\eta(\ell'\delta'-\ell_\gamma\delta_k)},~~~
\Delta\ell=\ell-\ell'-\ell_\gamma.
\end{eqnarray}
The second source of $\varphi$ enters through the variable $\tau_\rho$. For a circularly polarized laser $\tau_\rho$ appears as a phase $\exp(in\tau_\rho)$, while for a linearly polarized laser it appears as an argument $\gamma^\rmL=\sqrt{2}\gamma^\rmC\cos\tau_\rho$ of the Bessel function. Note that $\gamma^\rmC\propto\rho$ does not involve $\varphi$. Compute 
\begin{subequations}
\begin{eqnarray}
&&\rho^2
=\frac{k_+^2p_\perp^2}{p_+^4}
+\frac{k_\perp^2}{p_+^2}
-\frac{k_+(p_\perp^2+k_\perp^2-p_\perp^{\prime 2})}{p_+^3},
\\
&&\rho\cos\tau_\rho
=\rho\cos(\varphi-\eta\varphi_0),~~~
\rho\sin\tau_\rho=\rho\sin(\varphi-\eta\varphi_0),
\end{eqnarray}
\end{subequations}
i.e., $\tau_\rho=\varphi-\eta\varphi_0$, where $\varphi_0$ is determined by 
\begin{eqnarray}
&&\frac{k_+p_\perp}{p_+^2}
-\frac{k_\perp\cos\delta_k}{p_+}=\rho\cos\varphi_0,~~~
\frac{k_\perp}{p_+}\sin\delta_k
=\rho\sin\varphi_0.
\label{eq:varphi_0}
\end{eqnarray}
Finally, $\varphi$ appears as a polynomial of triangular functions in the $B_{0,1}$ functions, which involve $\varphi$ through the following variables: 
\begin{subequations}
\begin{eqnarray}
&&
r_\lambda=\frac{k_\perp}{k_+}e^{-i\lambda(\varphi+\eta\delta_k)},~~~
p_\sigma=p_\perp e^{i\sigma\varphi},~~~
\\
&&p'_\sigma=p'_\perp e^{i\sigma(\varphi-\eta\delta')},~~~
p'_{\sigmabar}=p'_\perp e^{-i\sigma(\varphi-\eta\delta')}.
\end{eqnarray}
\end{subequations}
Note that $R^3=\eta p_\perp p'_\perp\sin\delta'$ and $2\pvec'_\perp\cdot\pvec_\perp=p_\perp^2+p_\perp^{\prime 2}-k_\perp^2$ do not depend on azimuthal angles. 

Now we collect the above sources of $\varphi$ dependence and finish the $\varphi$ integrals. For a circularly polarized laser, this is trivial and yields Kronecker $\delta$'s involving various combinations of $\Delta\ell$, $n$, $\lambda$, $\sigma$, and $\sigma'=\pm\sigma$. This reflects the axial symmetry of the process around the laser propagation direction. Suppressing the following factors on the right hand side, 
\begin{eqnarray}
i^{-\Delta\ell}e^{i\eta(\ell'\delta'-\ell_\gamma\delta_k)}
\sqrt{\frac{2\pi}{\kappa}\frac{2\pi}{\kappa'}\frac{2\pi}{\kappa_\gamma}}
\delta(p_\perp-\kappa)\delta(p'_\perp-\kappa')\delta(k_\perp-\kappa_\gamma)
\big[2(E+M)(E'+M)\big]^{-1/2},
\label{eq:suppressed_factors}
\end{eqnarray}
we have 
\begin{eqnarray}
&&\int d\varphi~F^\rmC(\varphi,\varphi'_\eta,\varphi_{k\eta})
\nonumber
\\
&=&\int d\varphi~e^{i\Delta\ell\varphi}
\Big[B_0J_n(\gamma^\rmC)e^{in\tau_\rho}
+\frac{M\xi}{\sqrt{2}}B_1J_{n-\lambda}(\gamma^\rmC)e^{i(n-\lambda)\tau_\rho}\Big]
\nonumber
\\
&=&(0^\rmC_\textrm{s})J_n(\gamma^\rmC) 
+(1^\rmC_\textrm{s})\frac{M\xi}{\sqrt{2}}J_{n-\lambda}(\gamma^\rmC),
\label{eq:circular vortex amplitude}
\end{eqnarray}
where the subscript $\textrm{s}=\textrm{p, f}$ denotes the electron spin-preserved ($\sigma'=\sigma$) and -flipped ($\sigma'=\sigmabar=-\sigma$) cases. For the spin-preserved case, 
\begin{subequations}
\begin{eqnarray}
(0^\rmC_\textrm{p})
&=&(0_\textrm{p})
e^{i(\Delta\ell-\lambda)\eta\varphi_0}\delta_{\Delta\ell+n-\lambda,0},
\\
(1^\rmC_\textrm{p})
&=&(1_\textrm{p})
e^{i\Delta\ell\eta\varphi_0}\delta_{\Delta\ell+n-\lambda,0},
\end{eqnarray}
\end{subequations}
where 
\begin{subequations}
\begin{eqnarray}
\label{eq:circular vortex amplitude details}
(0_\textrm{p})
&=&\Big[(p'_+ +M)(p_+ +M)
+(\pvec'_\perp\cdot\pvec_\perp+iR^3\sigma)\Big]
\frac{k_\perp}{k_+}e^{-i\lambda\eta\delta_k}
\nonumber
\\
&&
-\left[2\delta_{\lambda\sigmabar}(E'+M)p_\perp
+2\delta_{\lambda\sigma}(E+M)p'_\perp e^{i\lambda\eta\delta'}\right],
\\
(1_\textrm{p})&=&+\bigg(\frac{\delta_{\lambda\sigma}}{p'_+}
+\frac{\delta_{\lambda\sigmabar}}{p_+}\bigg)(p'_+ +M)(p_+ +M)
\nonumber
\\
&&+\bigg(\frac{\delta_{\lambda\sigmabar}}{p'_+}
+\frac{\delta_{\lambda\sigma}}{p_+}\bigg)(\pvec'_\perp\cdot\pvec_\perp+i\sigma R^3),
\label{eq:circular vortex amplitude details additional}
\end{eqnarray}
\end{subequations}
and for the spin-flipped case, 
\begin{subequations}
\begin{eqnarray}
(0^\rmC_\textrm{f})
&=&
(0_\textrm{f})
e^{in(-\eta\varphi_0)}\delta_{\Delta\ell+n-\lambda+\sigma,0},
\\
(1^\rmC_\textrm{f})
&=&(1_\textrm{f})
e^{i(\Delta\ell+\sigma)\eta\varphi_0}\delta_{\Delta\ell+n-\lambda+\sigma,0},
\end{eqnarray}
\end{subequations}
where 
\begin{subequations}
\begin{eqnarray}
\label{eq:circular vortex amplitude details flipped}
(0_\textrm{f})
&=&\sigma\bigg\{\frac{k_\perp}{k_+}
\Big[(p_+ +M)p'_\perp e^{-i\sigma\eta\delta'}-(p'_+ +M)p_\perp\Big]
e^{-i\lambda\eta\delta_k}
\nonumber
\\
&&+2\delta_{\lambda\sigma}\Big[(p_+ +M)p^{\prime 3}-(p'_+ +M)p^3\Big]\bigg\},
\\
(1_\textrm{f})
&=&\sigma\bigg\{\bigg(\frac{\delta_{\lambda\sigma}}{p'_+}
+\frac{\delta_{\lambda\sigmabar}}{p_+}\bigg)(p_+ +M)p'_\perp
e^{-i\sigma\eta\delta'}
\nonumber
\\
&&-\bigg(\frac{\delta_{\lambda\sigmabar}}{p'_+}
+\frac{\delta_{\lambda\sigma}}{p_+}\bigg)(p'_+ +M)p_\perp\bigg\}.
\label{eq:circular vortex amplitude details flipped additional}
\end{eqnarray}
\end{subequations}

The $\varphi$ integrals are more involved in the case of a linearly polarized laser in which the argument of the Bessel function in integrands contains $\varphi$. But they are still integrable since only a single Bessel function is multiplied by an integral phase of $\varphi$. We reserve the calculation details of the basic integral $G$ in Appendix \ref{sec:appendix_C}. Suppressing again the same factors as in Eq.~\eqref{eq:suppressed_factors}, we have 
\begin{eqnarray}
&&\int d\varphi~F^\rmL(\varphi,\varphi'_\eta,\varphi_{k\eta})
\nonumber
\\
&=&\sum_{\nu=-\infty}^\infty J_\nu(\beta)
\int d\varphi~e^{i\Delta\ell\varphi}
\bigg[B_0J_{2\nu+n}(\gamma^\textrm{L})
+\frac{M\xi}{2}B_1\Big(J_{2\nu+n-1}(\gamma^\textrm{L}) +J_{2\nu+n+1}(\gamma^\textrm{L})\Big)\bigg]
\nonumber
\\
&=&\sum_{\nu=-\infty}^\infty 
\bigg[(0^\rmL_\textrm{s})+\frac{M\xi}{2}(1^\rmL_\textrm{s})\bigg]
J_\nu(\beta),
\label{eq:linear vortex amplitude}
\end{eqnarray}
where for the electron spin-preserved case, 
\begin{subequations}
\begin{eqnarray}
\label{linear amplitude 1}
(0^\rmL_\textrm{p})&=&(0_\textrm{p})G_{2\nu+n}^{\Delta\ell-\lambda}(\eta),
\\
(1^\rmL_\textrm{p})&=&
(1_\textrm{p})\Big[G_{2\nu+n-1}^{\Delta\ell}(\eta)
+G_{2\nu+n+1}^{\Delta\ell}(\eta)\Big],
\label{linear amplitude 2}
\end{eqnarray}
\end{subequations}
and for the spin-flipped case, 
\begin{subequations}
\begin{eqnarray}
(0^\rmL_\textrm{f})&=&(0_\textrm{f})G_{2\nu+n}^{\Delta\ell+\sigma-\lambda}(\eta),
\\
(1^\rmL_\textrm{f})
&=&(1_\textrm{f})
\Big[G_{2\nu+n-1}^{\Delta\ell+\sigma}(\eta)
+G_{2\nu+n+1}^{\Delta\ell+\sigma}(\eta)\Big].
\end{eqnarray}
\end{subequations}
Since nonvanishing of $G_\mu^m(\eta)$ demands the same parity of the integers $\mu,~m$, the following integers are even in their respective cases: $n+\Delta\ell-\lambda$ for $(0^\rmL_\textrm{p})$, $n+\Delta\ell \pm 1$ for $(1^\rmL_\textrm{p})$, $n+\Delta\ell+\sigma-\lambda$ for $(0^\rmL_\textrm{f})$, and $n+\Delta\ell +\sigma\pm 1$ for $(1^\rmL_\textrm{f})$. Since $\lambda=\pm 1$ and $\sigma=\pm 1$, this translates into the statement that $n$ has opposite parity to $\Delta\ell$ for the electron spin-preserved case and same parity as $\Delta\ell$ for the spin-flipped case.

\subsubsection{Bessel-vortex amplitude}
\label{sec:amplitude_bessel}

For the bookkeeping purpose we write down the scattering matrix element by collecting all factors except for the normalization factors of the initial and final states: 
\begin{eqnarray}
S_\B^\rmP
&=&(-ie)i^{-\Delta\ell}
\sqrt{\frac{\kappa}{2\pi}\frac{\kappa'}{2\pi}\frac{\kappa_\gamma}{2\pi}}
\frac{1}{\sqrt{2(E+M)(E'+M)}}\frac{1}{2\Delta}
(2\pi)\delta(P_+)
\sum_{n=n_0}^\infty 2\delta(u_n\Omega_- +P_-)\Upsilon_n^\rmP,
\label{eq:amplitude_bessel}
\end{eqnarray}
where 
\begin{eqnarray}
\Upsilon_n^\rmP
&=&\sum_{\eta=\pm}e^{i\eta(\ell'\delta'-\ell_\gamma\delta_k)}
\left\{
\begin{array}{ll}
\displaystyle (0^\rmC_\textrm{s})J_n(\gamma^\rmC) 
+(1^\rmC_\textrm{s})\frac{M\xi}{\sqrt{2}}J_{n-\lambda}(\gamma^\rmC), 
&\rmP=\rmC
\\
\displaystyle\sum_{\nu=-\infty}^\infty 
\bigg[(0^\rmL_\textrm{s})+\frac{M\xi}{2}(1^\rmL_\textrm{s})\bigg]
J_\nu(\beta), & \rmP=\rmL
\end{array}\right.,
\label{eq:upsilonP}
\end{eqnarray}
and $p_\perp=\kappa,~p'_\perp=\kappa',~k_\perp=\kappa_\gamma$ are implied on the right-hand side. We recall some features related to the laser polarization. For the circular polarization, only one Bessel function is involved in each term, and for given $\Delta\ell,~\sigma,~\sigma',~\lambda$ only one $n$ actually contributes ($n=\lambda-\Delta\ell>0$ for the electron spin-preserved case and $n=\lambda-\Delta\ell-\sigma>0$ for the electron spin-flipped case). This is again a result of angular momentum conservation in the $z$ direction. The initial electron has angular momentum component in the $z$ direction, $\ell+\sigma/2$, and the absorbed $n>0$ laser photons contribute $n$; the final electron has $\ell'+\sigma'/2$, and the radiated photon has $\ell_\gamma+\lambda$. Conservation of the angular momentum component gives $\Delta\ell+n=\lambda+(\sigma'-\sigma)/2$ which yields the Kronecker $\delta$. Since $n>0$, $\Delta\ell$ tends to be negative. In the case of the linear polarization without the axial symmetry, each term is a product of three Bessel functions, and there are double infinite sums over $n$ and $\nu$. Considering its simplicity our numerical analysis will be focused mainly on the case of a circularly polarized laser.

\subsection{Amplitude squared and decay rate for vortex electrons and photon}
\label{sec:decay_Bessel}

\subsubsection{Case for Bessel vortex}
\label{sec:rate_bessel}

On taking the absolute square of $S_\B^\rmP$ in Eq.~\eqref{eq:amplitude_bessel}, we first cope with the sum over $n$ and squaring of the $\delta$ functions. Given a point in the final-state phase space of $\pvec',~\kvec'$, the argument of $\delta(u_n\Omega_- +P_-)$ being linear in $n$ can vanish for at most one value of $n$. There is thus no interference between different $n$ terms. This is consistent with the physical interpretation that $n$ may be understood as the net number of laser photons involved in the process. Now consider the square of $\delta$ functions. The one involving the conserved plus component of momentum is, 
\begin{eqnarray}
&&2\pi\delta(P_+)=\int \frac{1}{2}dx_-~\exp\Big(\frac{i}{2}x_-P_+\Big)
\Longrightarrow
\big[2\pi\delta(P_+)\big]^2
= 2\pi\delta(P_+)\frac{L_-}{2}.
\end{eqnarray}
More care must be practiced for $[\delta(u_n\Omega_- +P_-)]^2$. The variable $u$ was introduced as an auxiliary parameter to uniformize momentum conservation among all components in the presence of a plane-wave laser. To proceed in a clear manner, we go back to the original $u$ and $\tau$ integrals in \cref{sec:phases}. Finishing first the $\tau$ ($u$) integral and then the $u$ ($\tau$) integral yields the first (second) equality respectively: 
\begin{eqnarray}
&&
\int d\tau\int du~\delta(u\Omega_- +P_-)e^{i(u+2\beta-n)\tau}
\nonumber
\\
&=&2\pi\delta(u_n\Omega_- +P_-)
\nonumber
\\
&=&\frac{1}{\Omega_-}2\pi\delta(u+2\beta-n)\Big|_{u=-P_-/\Omega_-}.
\end{eqnarray}
Squaring the above gives 
\begin{eqnarray}
&&\big[2\pi\delta(u_n\Omega_- +P_-)\big]^2
=\frac{1}{\Omega_-^2}\big[2\pi\delta(u+2\beta-n)|_{u=-P_-/\Omega_-}]^2
\nonumber
\\
&&=\frac{1}{\Omega_-^2}2\pi\delta(u+2\beta-n)|_{u=-P_-/\Omega_-}\int d\tau~e^{i0\cdot\tau}
=2\pi\delta(u_n\Omega_-+P_-)\frac{L_+}{2}.
\end{eqnarray}
The amplitude squared is thus 
\begin{eqnarray}
|S_\B^\rmP|^2
&=&e^2\frac{\kappa}{2\pi}\frac{\kappa'}{2\pi}
\frac{\kappa_\gamma}{2\pi}
\frac{1}{2(E+M)(E'+M)}\frac{1}{(2\Delta)^2}
2\pi\delta(P_+)\frac{L_-}{2}\frac{L_+}{2}\frac{2}{\pi}
\sum_{n=n_0}^\infty \delta(u_n\Omega_- +P_-)|\Upsilon_n^\rmP|^2.
\end{eqnarray}
Multiplying $|S_\B^\rmP|^2$ by normalization factors and number densities of final-state particles and dividing it by the time $L_+$, we obtain the decay rate \cite{Kotkin:1992bj}: 
\begin{eqnarray}
d\Gamma^\rmP_\B
&=&\frac{|S_\B^\rmP|^2}{L_+}(N_p^\B)^2 
\frac{d\kappa' dp'_+}{2p'_+2\pi}\frac{d\kappa_\gamma dk_+}{2k_+2\pi}
\nonumber
\\
&=&e^2\frac{\kappa}{2\pi}\frac{\kappa'}{2\pi}\frac{\kappa_\gamma}{2\pi}
\frac{1}{2(E+M)(E'+M)} 
\frac{1}{2\Delta^2}\frac{\pi}{R}\frac{1}{2p_+}\delta(P_+)
\sum_{n=n_0}^\infty \delta(u_n\Omega_-+P_-)|\Upsilon_n^\rmP|^2
\frac{d\kappa' dp'_+}{2p'_+2\pi}\frac{d\kappa_\gamma dk_+}{2k_+2\pi}.
\label{eq:rate_bessel}
\end{eqnarray}
The presence of the awkward factor $1/R$ in the above signifies the nonnormalizability of the Bessel vortex in the transverse plane. Now we employ the Bessel-Gaussian vortex to normalize the initial electron. 

\subsubsection{Modifications for an initial Bessel-Gaussian vortex electron}
\label{sec:rate_gauss}

The above result for the decay rate is singular at $\Delta=0$ when the triangle formed by the three transverse momenta $\pvec_\perp=\pvec'_\perp+\kvec_\perp$ shrinks to a line. This singularity of order $\Delta^{-2}$ in the rate is not integrable \cite{Ivanov:2011kk,Ivanov:2011bv}. Now we show that it can be made integrable by employing a Gaussian-smeared Bessel-Volkov vortex for the initial electron, so that the rate becomes a measurable quantity. Since such a vortex is a linear composition of the Bessel-Volkov solution to the Dirac equation which by itself is also linear, it is still a solution. But it is not unique to compose the solution. As we argued previously, it is more natural to employ light-front coordinates in the presence of a plane-wave laser. We therefore propose to work with a Gaussian vortex of fixed plus momentum, Eq.~\eqref{eq:initial_eBG}.  

There are two modifications to the first equality for decay rate in Eq.~\eqref{eq:rate_bessel}. First, the normalization factor $N_p^\B$ for the initial electron should be replaced by $N_p^\BG$ in Eq.~\eqref{eq:nor_gauss}, which effectively removes the factor $\pi/R$ while keeping $1/(2p_+)$ in the second equality of Eq.~\eqref{eq:rate_bessel}. Second, the $S$ matrix element $S_\B^\rmP$ in  Eq.~\eqref{eq:amplitude_bessel} should be weighted by the Gaussian function in Eq.~\eqref{eq:gaussian} and integrated over $\kappa$ to become the $S$ matrix element $S_\BG^\rmP$, as we detail below. 

Let us enumerate all $\kappa$ dependence other than the Gaussian function in $S_\B^\rmP$. First, the trivial square root factors. Second, $\delta(u_n\Omega_-+P_-)$ depends on $\kappa$ through $p_-$ in $P_-$. Since $p_+\delta p_--2\kappa\delta\kappa=0$ for an on-shell electron with fixed $p_+$, $\delta p_-\ll\delta\kappa$ for $\kappa\ll p_+$ which is the vortex that can be practically realized. Note $\beta$ in $u_n=n-2\beta$ is independent of $\kappa$, see Eq.~\eqref{eq:beta_gammaC}. That there are two solutions to $n$ on varying $\kappa$ requires at least $\kappa\delta\kappa\ge\omega p_+$. Considering further the Gaussian suppression for $\kappa$ deviating from its central value, the effect due to $\kappa$ variation must be tiny. Neglecting it there will be no interference between different $n$ terms when squaring $S_\B^\rmP$ as is the case with pure Bessel vortex states. The third source of $\kappa$ dependence is $\Upsilon_n^\rmP$, Eq.~\eqref{eq:upsilonP}. The good news is that the argument $\gamma^\rmC$ in Eq.~\eqref{eq:beta_gammaC} of the Bessel function in the case of a circularly polarized laser is independent of $\kappa$. This is because the magnitude $\rho$ is actually independent of transverse momenta although its definition as a vector in Eq.~\eqref{eq:rho} involves them. We verify this as follows. By definition, 
\begin{eqnarray}
\vec\rho^2=\frac{k_+^2}{p_+^4}p_\perp^2 +\frac{k_\perp^2}{p_+^2} 
-\frac{k_+}{p_+^3}2\kvec_\perp\cdot\pvec_\perp.
\end{eqnarray}
Using $2k\cdot p=k_+p_- +k_-p_+ -2\kvec_\perp\cdot\pvec_\perp$ and on-shell conditions $p_\perp^2=p_+p_- -M^2$ and $k_\perp^2=k_+k_-$ to replace all scalar products of transverse momenta, the above becomes
\begin{eqnarray}
\vec\rho^2
=\frac{k_+}{p_+^3}2k\cdot p -\frac{k_+^2}{p_+^4}M^2.
\end{eqnarray}
Identifying $u$ in Eq.~\eqref{eq:u} with $u_n$ through the $\delta$ function, we have $2k\cdot p=u_n2p'_+\omega$, so that 
\begin{eqnarray}
\vec\rho^2=\frac{k_+}{p_+^3}u_n2p'_+\omega -\frac{k_+^2}{p_+^4}M^2, 
\end{eqnarray}
which is independent of transverse momenta as claimed. In the case of a linearly polarized laser, as we show in Appendix \ref{sec:appendix_C} this only brings in integral phases of $\varphi_0$ as in the case of circular polarization. The remaining $\kappa$ dependence in $\Upsilon_n^\rmP$ appears in the form of polynomials in $\kappa,~E,~p_-$ and in the form of integer phases of $\delta',~\delta_k$. 

All $\kappa$ dependence mentioned above is smooth. The only apparently singular behavior enters through the factor $\Delta^{-1}$, which we now turn to discuss. To simplify notations, we will use the variables: 
\begin{eqnarray}
x=\kappa,~b=\kappa',~c=\kappa_\gamma.
\end{eqnarray}
Given a point in the final-state phase space, that transverse momenta form a triangle demands 
\begin{eqnarray}
x_1\le x\le x_2,~~~x_1=|b-c|,~~~x_2=b+c. 
\end{eqnarray}
The factor $\Delta^{-1}$ introduces apparent singularities at the delimiters of the integral over $\kappa=x$: 
\begin{eqnarray}
\Delta
=\frac{1}{4}\sqrt{(x+x_1)(x+x_2)}\sqrt{(x_2-x)(x-x_1)}.
\end{eqnarray}
The scattering matrix element $S_\B^\rmP$ in Eq.~\eqref{eq:amplitude_bessel} is replaced by 
\begin{eqnarray}
S_\BG^\rmP&=&(-ie)i^{-\Delta\ell}
(2\pi)\delta(P_+)\sum_{n=n_0}^\infty 2\delta(u_n\Omega_- +P_-)
I^\rmP_\BG,
\label{eq:amplitude_BG}
\end{eqnarray}
where 
\begin{subequations}
\begin{eqnarray}\label{BG integral}
I^\rmP_\BG&=&\int_{x_1}^{x_2}dx \frac{g(x)}{\sqrt{(x_2-x)(x-x_1)}},
\\
g(x)
&=&\frac{2f(x;\kappa_0,\sigma_\perp)}{\sqrt{(x+x_1)(x+x_2)}}
\bigg[\sqrt{\frac{\kappa}{2\pi}\frac{\kappa'}{2\pi}\frac{\kappa_\gamma}{2\pi}}
\frac{e^{i\eta(\ell'\delta'-\ell_\gamma\delta_k)}}{\sqrt{2(E+M)(E'+M)}}
\Upsilon_n^\rmP\bigg]_{\kappa=x}.
\end{eqnarray}
\end{subequations}
We have set $\kappa$ in $\delta(u_n\Omega_-+P_-)$ to its central value $\kappa_0$, i.e., $p_-=(M^2+\kappa_0^2)/p_+$, and included in the function $g(x)$ all smooth dependence of $\kappa$ and the Gaussian function. The integral $I_\BG$ is manipulated in Appendix \ref{sec:appendix_D} into a form that involves no apparent singularities and is amenable to numerical computation. 
The decay rate is, 
\begin{eqnarray}
\label{explicit decay width}
d\Gamma^\rmP_\BG
&=&\frac{|S_\BG^\rmP|^2}{L_+}(N_p^\BG)^2 
\frac{d\kappa' dp'_+}{2p'_+ 2\pi}\frac{d\kappa_\gamma dk_+}{2k_+2\pi}
\nonumber
\\
&=&e^22\delta(P_+)\sum_{n=n_0}^\infty \delta(u_n\Omega_-+P_-)
|I^\rmP_\BG|^2 \frac{1}{2p_+}
\frac{d\kappa' dp'_+}{2p'_+ 2\pi}\frac{d\kappa_\gamma dk_+}{2k_+2\pi}.
\end{eqnarray}
The $\kappa'$ and $p'_+$ integrals are finished by the two $\delta$ functions, i.e., 
\begin{subequations}
\begin{eqnarray}
p'_+&=&
E(1+v\cos\theta_p)-\omega_\gamma(1+\cos\theta_\gamma),
\\
\kappa'^2
&=&
\kappa^2+\omega^{2}_\gamma\sin^{2}\theta_\gamma
+2n\omega\left[E(1+v\cos\theta_p)-\omega_\gamma(1+\cos\theta_\gamma)\right]
\nonumber
\\
&&
-2\omega E(1-v\cos\theta_p\cos\theta_\gamma)
-\frac{1}{2}M^2\xi^2\frac{\omega_\gamma(1+\cos\theta_\gamma)}{E(1+v\cos\theta_p)},
\end{eqnarray}
\end{subequations}
where $v$ is the magnitude of the velocity of the initial electron. For the emitted vortex photon, we employ its angular frequency $\omega_\gamma$ and opening angle $\theta_\gamma$ to characterize its kinematical distributions by the relation, $dk_+d\kappa_\gamma=\omega_\gamma(1+\cos\theta_\gamma)d\omega_\gamma d\theta_\gamma$. With all of the above manipulations, the two-dimensional differential decay rate in terms of the emitted photon's variables becomes,
\begin{eqnarray}
d\Gamma^\rmP_\BG
&=&
\frac{e^2}{32\pi^2 E(1+v\cos\theta_p)} \sum_{n=n_0}^\infty \frac{|I^\rmP_\BG|^2}{\kappa'}d\omega_\gamma d\theta_\gamma.
\end{eqnarray}
Given the variables $(E,\theta_p,v)$ of the initial electron and the laser's angular frequency $\omega$ and parameter $\xi$, the angular frequency of the emitted vortex photon is bound by $\omega_-\le\omega_\gamma\le\omega_+$ for a given $\theta_\gamma$, where 
\begin{eqnarray}
\label{eq:photon energy bounds}
\omega_{\pm} &=& \frac{n\omega E(1+v\cos\theta_{p})}
{E[1-v\cos(\theta_\gamma\mp\theta_p)]+\big\{n\omega
+M^2\xi^2/[4E(1+v\cos\theta_{p})]\big\}(1+\cos\theta_\gamma)}.
\end{eqnarray}
Note that $\omega_\pm$ coincide at $\theta_\gamma=\pi$ and are equal to $n\omega$.

\section{Numerical analysis}
\label{sec:numerical}

In this section we evaluate numerically the rate for all-vortex nonlinear Compton scattering based on the analytical result in \cref{sec:analytical}. We will focus mainly on the case of a circularly polarized laser, and discuss briefly the case of a linearly polarized laser. For the laser parameters, we assume its angular frequency $\omega=1~\eV$ and dimensionless intensity $\xi=0.1,~1,~5$. This corresponds to a laser intensity in the range from $2\times 10^{16}~\textrm{W/cm}^{2}$ to $5\times 10^{19}~\textrm{W/cm}^{2}$, which can be achieved in current experiments~\cite{Gonoskov:2021hwf}. We assume that the initial electron has an energy $E=1~\GeV$ which could be obtained by accelerating a low energy vortex electron to higher energy \cite{Karlovets:2020tlg}. Unless otherwise stated, we assume its Bessel-Gaussian vortex has a deviation $\sigma_\perp$ of one percent of its central value $\kappa_0$, $\sigma_\perp=0.01\kappa_0$. 
We recall that angular momentum is conserved in the longitudinal direction for a circularly polarized laser (see Eq.~\eqref{eq:circular vortex amplitude}), which results in the constraint $\Delta\ell+n-\lambda=0$ for the electron spin-preserved channel and $\Delta\ell+n-\lambda+\sigma=0$ for the electron spin-flipped channel. There is no such a conservation law in the case of a linearly polarized laser, hence we have to sum up the contributions from all possible harmonic numbers. 

\subsection{Circularly polarized laser with $\xi=0.1$}
\subsubsection{The electron spin-preserved case}

We start with the case when the electron preserves its spin $\sigma'=\sigma$ upon colliding against a less intense laser with $\xi=0.1$. The exchanges of angular momenta among the electron, photon, and laser are discussed, by analyzing a number of combinations of the OAMs of the initial vortex electron $\ell$, the final vortex electron $\ell'$, and the emitted vortex photon $\ell_\gamma$. The impact of the opening angle $\theta_p$ of the initial electron is also analyzed.

\begin{figure}[H]
    \centering
    \includegraphics[width=0.75\textwidth]{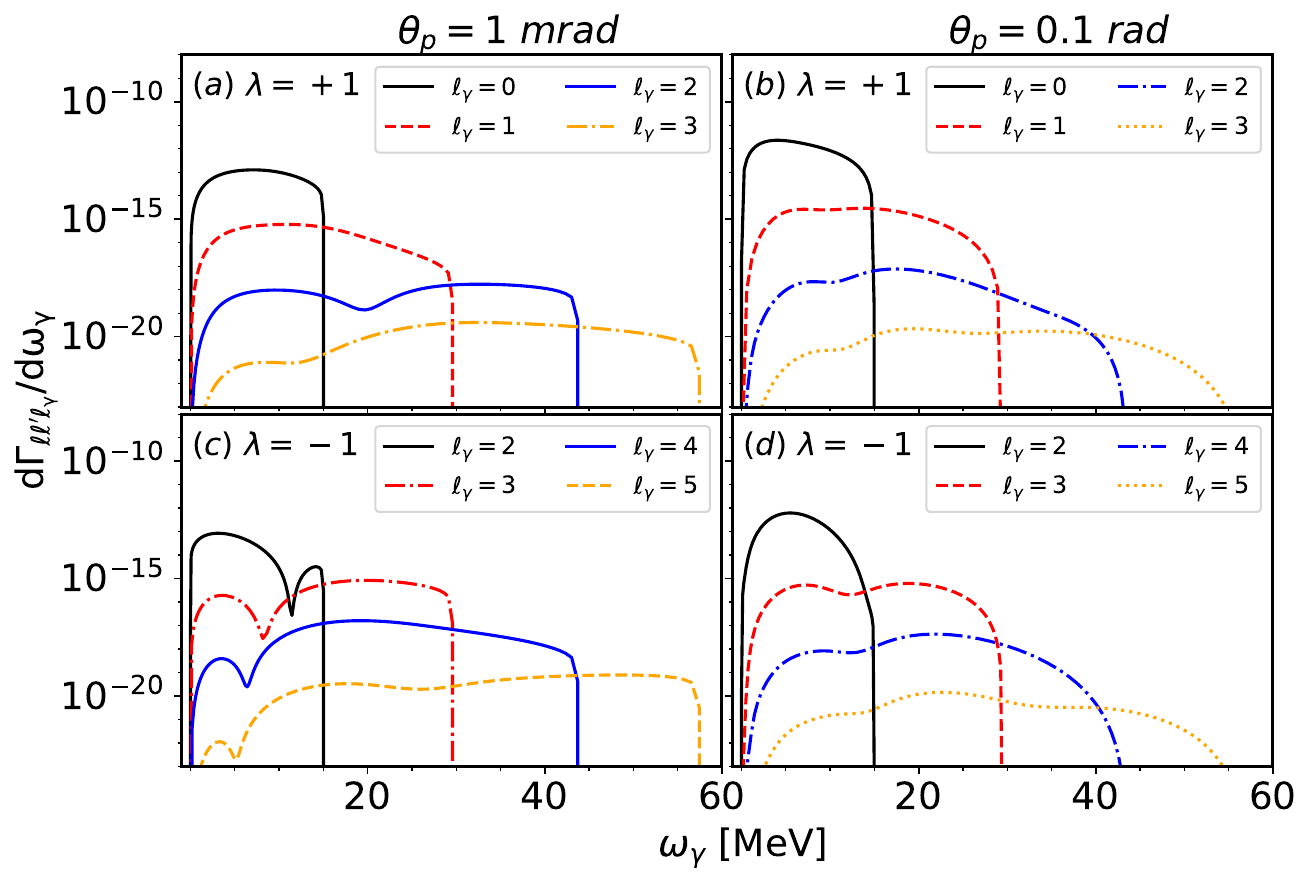}
    \caption{The distribution $d\Gamma_{\ell\ell'\ell_\gamma}/d\omega_\gamma$ of the emitted vortex $\gamma$ photon at $\theta_p = 1~\textrm{mrad}$ ($\theta_p = 0.1~\textrm{rad}$) in panels (a) and (c) ((b) and (d)), and for the photon helicity $\lambda=+1$ ($\lambda=-1$) in panels (a) and (b) ((c) and (d)). In all panels $\ell=\ell'=10$, $\sigma=\sigma'=+1$, and $\xi=0.1$.}
    \label{1d_xi0.1}
\end{figure}

In Fig.~\ref{1d_xi0.1} we depict the spectrum $d\Gamma_{\ell\ell'\ell_\gamma}/d\omega_\gamma$  
of the emitted vortex photon for the case of $\ell=\ell'=10$ and $\sigma=\sigma'=+1$. Our analysis shows that the case with $\sigma=\sigma'=-1$ is very similar. 
Two values of the opening angle of the initial electron $\theta_p=1~\textrm{mrad}$ and $\theta_p=0.1~\textrm{rad}$ are considered. The spectrum $d\Gamma_{\ell\ell'\ell_\gamma}/d\omega_\gamma$ is obtained by finishing the opening angle $\theta_\gamma$ integral in the two-dimensional distribution $d\Gamma_{\ell\ell'\ell_\gamma}/(d\omega_\gamma d\theta_\gamma$). 
The OAM $\ell_\gamma$ of the emitted photon is restricted by the angular momentum conservation $\ell_\gamma=\ell-\ell'+n-\lambda$, as discussed above. For a given $\ell_\gamma$, its energy lies between $n\omega$ and $\omega_{+}$, as can be clearly seen in Fig.~\ref{1d_xi0.1}. While the maximum energy of the emitted photon increases with its OAM $\ell_\gamma$, the rate $d\Gamma_{\ell\ell'\ell_\gamma}/d\omega_\gamma$ decreases by a few orders of magnitude. The combination of these two factors offers a chance to separate the emitted vortex $\gamma$ photons with different OAMs; namely, in every given energy range, one value of $\ell_\gamma$ dominates over other values. Furthermore, a double-peak or multi-peak structure can be observed in $d\Gamma_{\ell\ell'\ell_\gamma}/d\omega_\gamma$ for a large enough $\ell_\gamma$.

\begin{figure}[H]
    \centering
    \includegraphics[width=0.9\linewidth]{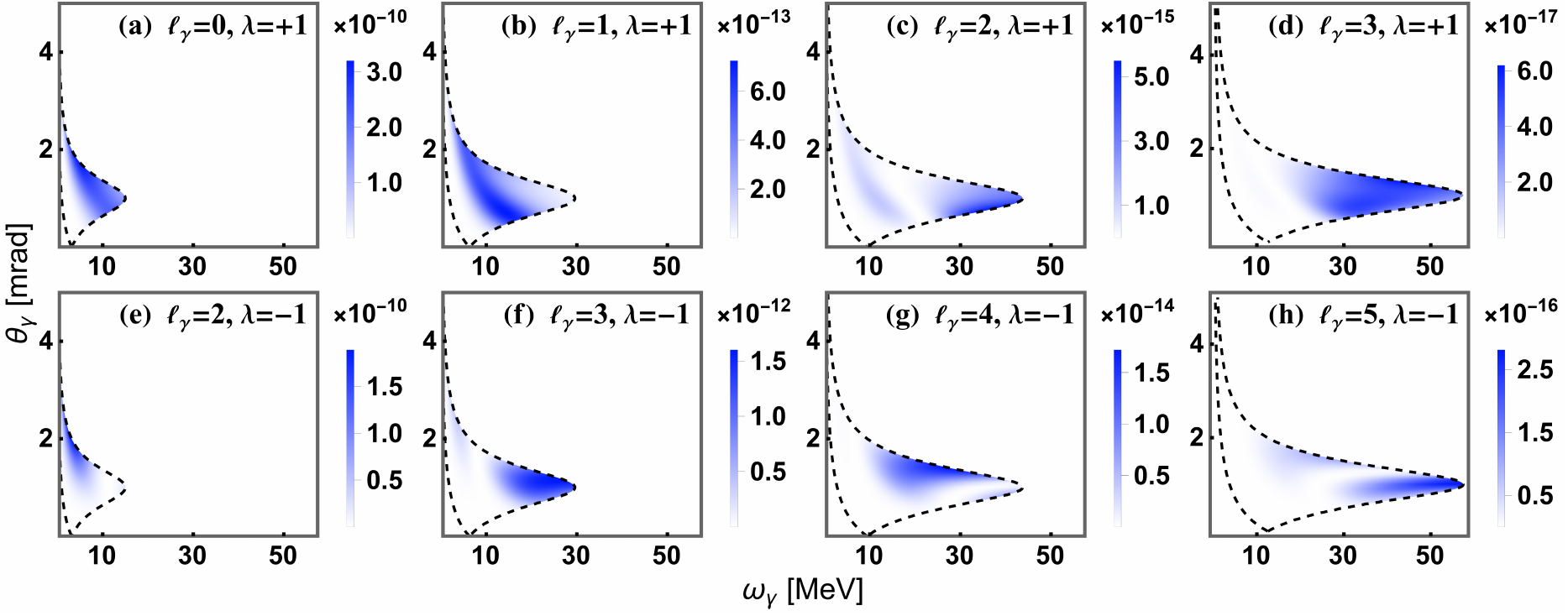}
    \caption{The two-dimensional distribution $d\Gamma_{\ell\ell'\ell_\gamma}/(d\omega_\gamma d\theta_\gamma$) for $\lambda=+1$ ($\lambda=-1$) in upper (lower) panels. In all panels $\ell=\ell'=10$, $\sigma=\sigma'=+1$, $\theta_p=1~\textrm{mrad}$, and $\xi=0.1$.}
    \label{2d_xi0.1_1mrad}
\end{figure}

The impact of the opening angle $\theta_p$ of the initial vortex electron which represents the effects of its transverse momentum is shown clearly in Fig.~\ref{1d_xi0.1}. Increasing $\theta_p$ results in a larger contribution from the terms in Eqs.~\eqref{eq:circular vortex amplitude details} and \eqref{eq:circular vortex amplitude details additional} proportional to transverse momenta in the amplitude. However, as the spin-preserved case is not very sensitive to transverse momenta, the difference in the rates between the two opening angles $\theta_p = 1~\textrm{mrad}$ and $\theta_p = 0.1~\textrm{rad}$ is not very large compared to the spin-flipped case to be studied later.

\begin{figure}[H]
    \centering
    \includegraphics[width=1\linewidth]{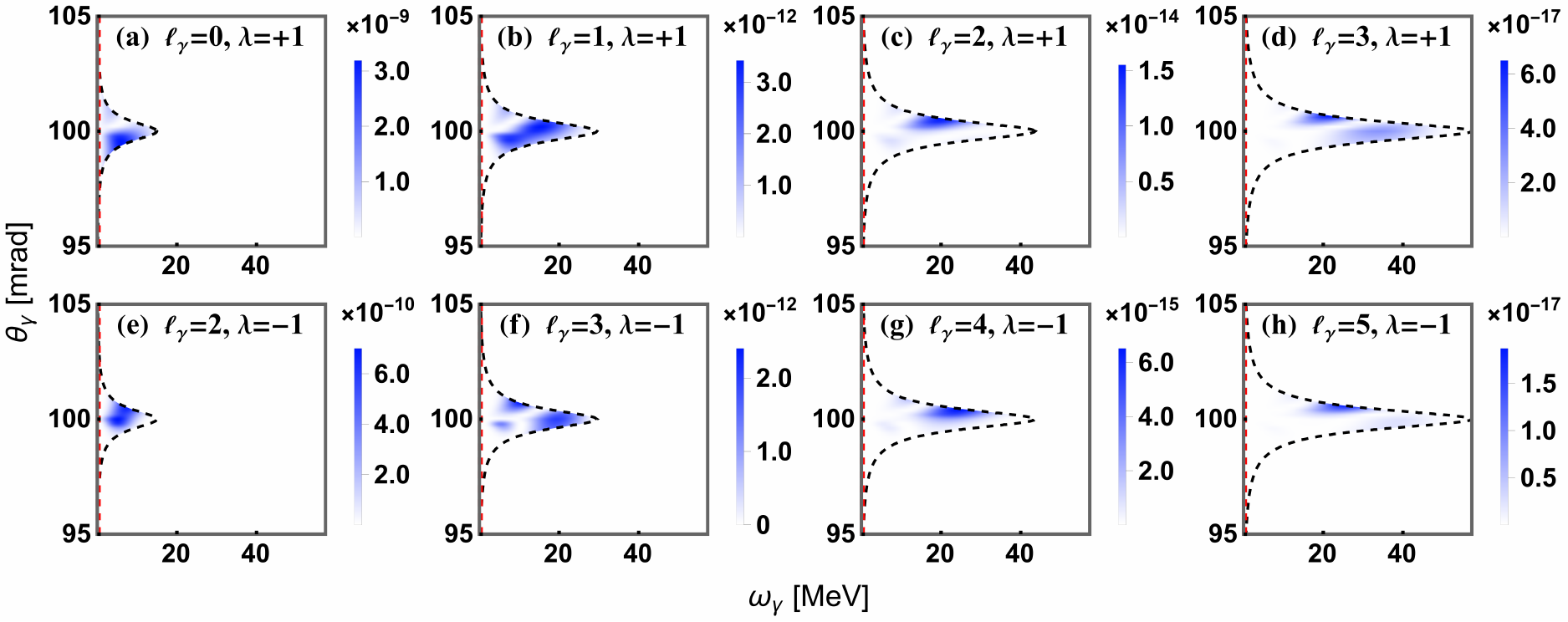}
    \caption{Same as Fig.~\ref{2d_xi0.1_1mrad} except for $\theta_p=0.1~\textrm{rad}$. 
    }
    \label{2d_xi0.1_0.1rad}
\end{figure}

\begin{figure}[H]
    \centering
    \includegraphics[width=0.75\linewidth]{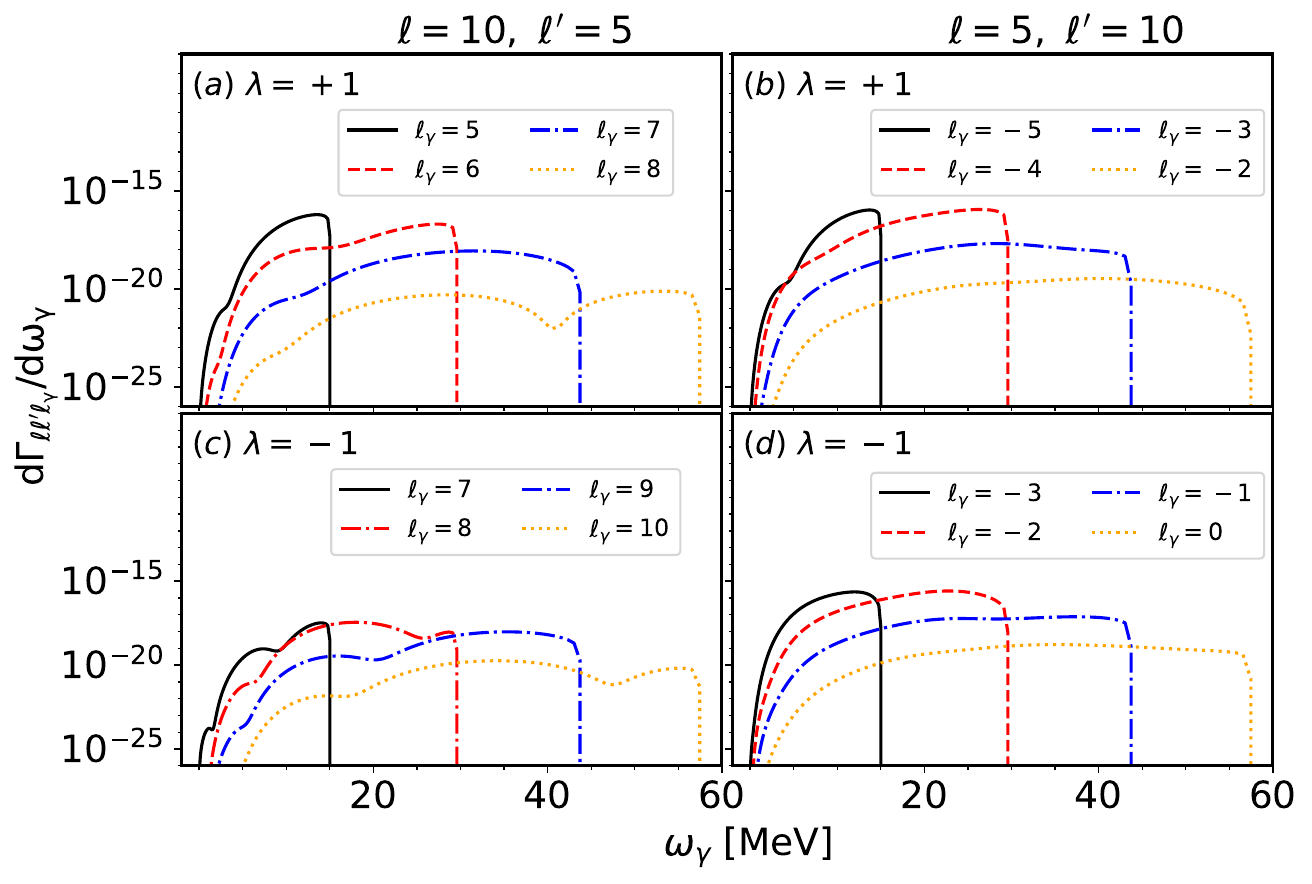}
    \caption{
    The distribution $d\Gamma_{\ell\ell'\ell_\gamma}/d\omega_\gamma$ of the emitted vortex $\gamma$ photon for $\ell=10,~\ell'=5$ ($\ell=5,~\ell'=10$) in panels (a) and (c) ((b) and (d)) and $\lambda=+1$ ($\lambda=-1$) in panels (a) and (b) ((c) and (d)), and for various $\ell_\gamma$. In all panels $\sigma=\sigma'=+1$, $\theta_p=1~\textrm{mrad}$, and $\xi=0.1$.}
    \label{1d_diff_ell}
\end{figure}

We further analyze the two-dimensional decay-rate distribution $d\Gamma_{\ell\ell'\ell_\gamma}/(d\omega_\gamma d\theta_\gamma)$. The results are presented in Fig.~\ref{2d_xi0.1_1mrad} for $\theta_p=1~\textrm{mrad}$ and in Fig.~\ref{2d_xi0.1_0.1rad} for $\theta_p=0.1~\textrm{rad}$. Again, the case of $\ell=\ell'=10$ and $\sigma=\sigma'=+1$ is considered here. The region bound by $\omega_\pm$ in Eq.~(\ref{eq:photon energy bounds}) is the one enclosed by a dashed curve. It can be observed in Figs.~\ref{2d_xi0.1_1mrad} and \ref{2d_xi0.1_0.1rad} that the opening angle $\theta_\gamma$ of the emitted vortex $\gamma$ photon locates mainly near the value of the opening angle $\theta_p$ of the initial vortex electron. For the case of a larger $\theta_p=0.1~\textrm{rad}$, $\omega_{-}$ is very small which we indicate in Fig.~\ref{2d_xi0.1_0.1rad} by a red dashed curve.

Now we turn to the case when the OAM $\ell'$ of the final electron is different from that of the initial electron $\ell$. The spectrum $d\Gamma_{\ell\ell'\ell_\gamma}/d\omega_\gamma$ of the emitted vortex $\gamma$ photon is presented in Fig.~\ref{1d_diff_ell} for the cases $(\ell,\ell')=(10,5),(5,10)$. Here we focus on the condition of $\sigma=\sigma'=+1$ and $\theta_p=1~\textrm{mrad}$. Compared with the results in Fig.~\ref{1d_xi0.1} with $\ell'=\ell$, the change in the rate with $\ell_\gamma$ is not as significant, although the strong $\ell_\gamma$-dependence of the maximum energy of the emitted photon still remains. In addition, the feature of $d\Gamma_{\ell\ell'\ell_\gamma}/d\omega_\gamma$ changes from a single-peak structure to a multi-peak structure with increasing $|\ell_\gamma|$. This is a combined effect of phases and Bessel functions in the scattering amplitude. For a small opening angle $\theta_p$ under consideration here, the photon's transverse momentum $k_\perp$ is much smaller than those of the initial and final electrons $p_\perp,~p'_\perp$ which thus stay close to each other. Then according to Eqs.~\eqref{angle between electron and electron} and \eqref{angle between electron and photon}, $\delta'$ is always small while $\delta_k$ varies in the whole range from 0 to $\pi$ making $\delta_k$-dependent phases oscillate rapidly. Those phases enter in the amplitude both as an $\ell_\gamma$-dependent one $e^{-i\ell_\gamma\delta_k}$ and as a helicity-dependent one $e^{-i\lambda\eta\delta_k}$. Furthermore, there are also $\delta_k$-dependent phases that are accompanied by Bessel functions, $e^{-in\eta\varphi_0}J_n(\gamma^\rmC)$ and $e^{-i(n-\lambda)\eta\varphi_0}J_n(\gamma^\rmC)$ where $\varphi_0$ depends on $\delta_k$ through Eq.~\eqref{eq:varphi_0}. The joint effect of all these causes the multi-peak structure for a large enough $\ell_\gamma$. This feature can be viewed as a result of scattering of particles carrying OAMs, and the multi-peak structure could help to identify the OAM and spin of the vortex particle. 

The comparison of Figs.~\ref{1d_diff_ell} and \ref{1d_xi0.1} also shows that the decay rate is suppressed with increasing $|\ell_\gamma|$ (or $|\ell-\ell'|$ for fixed $n,~\lambda$). 
When $|\ell-\ell'|$ or $|\ell_\gamma|$ is sufficiently large, the phase factor oscillates strongly. For Bessel-Gaussian states, this makes the amplitude oscillate strongly, which tends to diminish the integration in Eq.~(\ref{BG integral}). In fact, $|\ell-\ell'|$ (or $|\ell_\gamma|$) should not be too large as compared with $\kappa_\gamma/\sigma_\perp$~\cite{Ivanov:2011bv}, otherwise the decay rate would be very small. For the values of the parameters considered here, the photon transverse momentum $\kappa_\gamma$ is indeed not much larger than $\sigma_\perp$, and we therefore see from Fig.~\ref{1d_diff_ell} the start of the suppression.

\begin{figure}[H]
    \centering
    \includegraphics[width=0.75\linewidth]{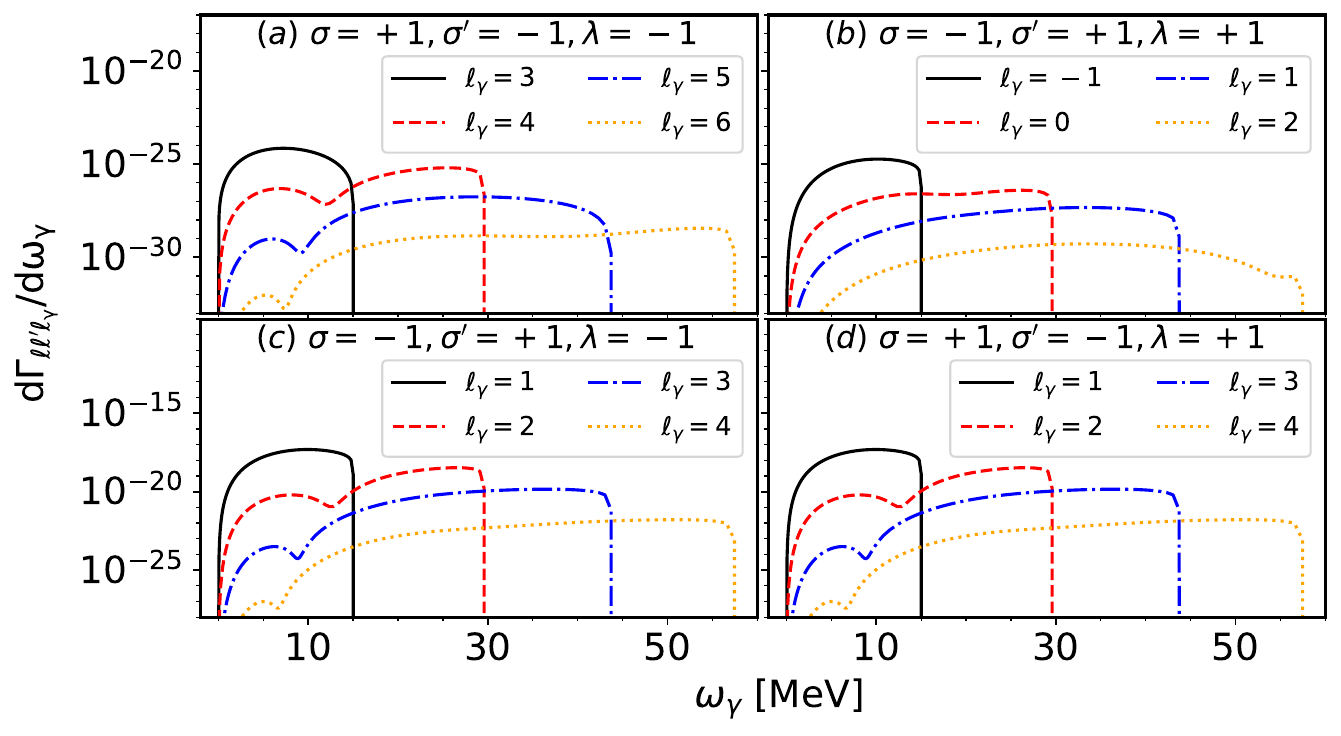}
    \caption{The distribution $d\Gamma_{\ell\ell'\ell_\gamma}/d\omega_\gamma$ of the emitted vortex $\gamma$ photon in the electron spin-flipped case: $\sigma'=-\sigma=-1$ ($\sigma'=-\sigma=+1$) in panels (a,d) (panels (b,c)) and $\lambda=-1$ ($\lambda=+1$) in panels (a,c) (panels (b,d)). In all panels, $\xi=0.1$, $\ell=\ell'=10$, and $\theta_p=1~\textrm{mrad}$.}
    \label{1d_flipped_1mrad}
\end{figure}

\subsubsection{The electron spin-flipped case}

\begin{figure}[H]
    \centering
    \includegraphics[width=0.75\linewidth]{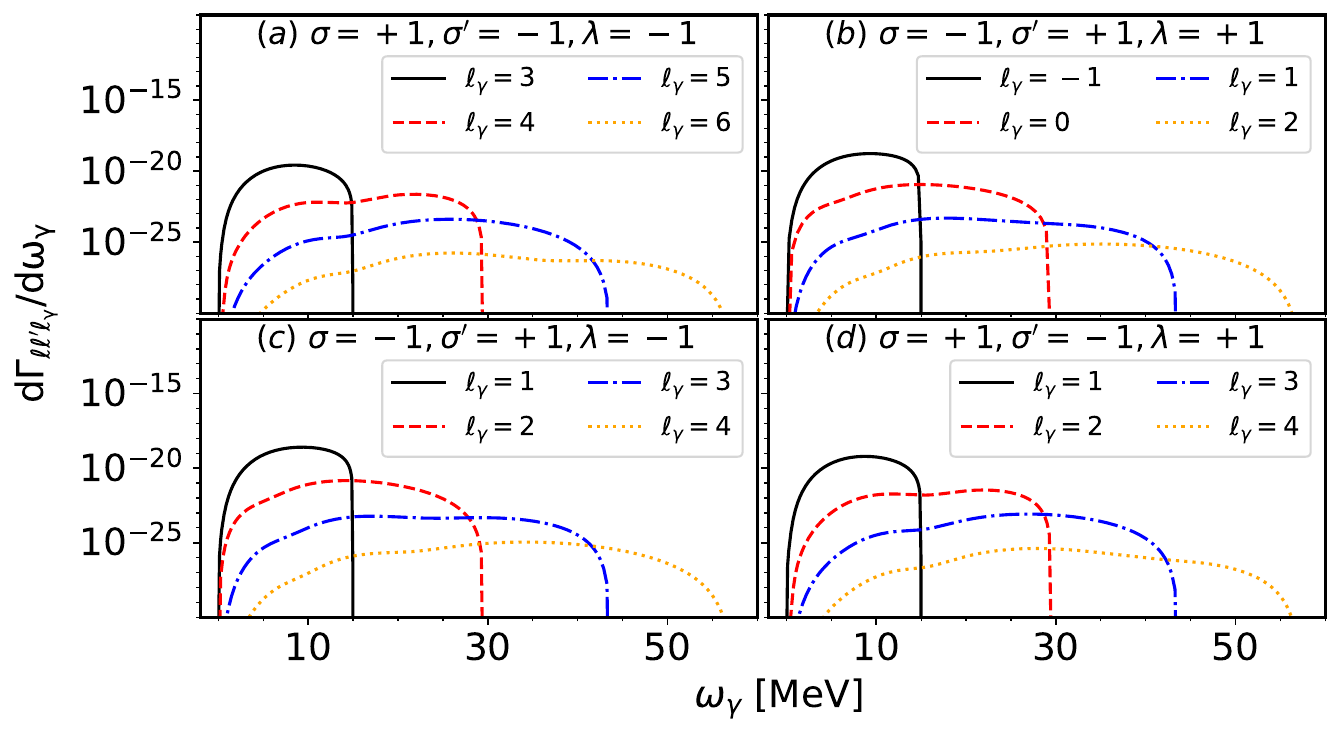}
    \caption{Same as Fig.~\ref{1d_flipped_1mrad} except for $\theta_p=0.1~\textrm{rad}$.
    }
    \label{1d_flipped_0.1rad}
\end{figure}

For the electron spin-flipped case with $\sigma'=-\sigma=\pm 1$, we focus on the case of $\ell=\ell'=10$, and present our results for the opening angle $\theta_p = 1~\textrm{mrad}$ of the initial vortex electron in Fig.~\ref{1d_flipped_1mrad}, and for $\theta_p = 0.1~\textrm{rad}$ in Fig.~\ref{1d_flipped_0.1rad}. Comparison with Fig.~\ref{1d_xi0.1} shows that the rate for the spin-flipped case is in general smaller by orders of magnitude than the one for the spin-preserved case. This indicates the difficulty in changing and manipulating the electron spin. As shown in Fig.~\ref{1d_flipped_1mrad}(c) and Fig.~\ref{1d_flipped_1mrad}(d), there is a symmetry between the cases $\sigma=\lambda=-1$ and $\sigma=\lambda=+1$ when $\theta_p$ is small. This is because of the same phase factor and the symmetric form of the dominant terms in the amplitude, Eqs.~\eqref{eq:circular vortex amplitude details flipped} and \eqref{eq:circular vortex amplitude details flipped additional}. We observe in Fig.~\ref{1d_flipped_1mrad} that the rate for the case $\sigma=-\lambda$ (panels (a) and (b)) is much smaller than the case $\sigma=\lambda$ (panels (c) and (d)). This is because the term $(p^{+}+M)p'^{3}-(p'^{+}+M)p^{3}$, which is much larger than other terms in the amplitude [see Eq.~\eqref{eq:circular vortex amplitude details flipped}] when the opening angle $\theta_p$ is small, only contributes for the case $\sigma=\lambda$. 

When the opening angle $\theta_p$ of the initial vortex electron is larger, the terms proportional to transverse momenta in the amplitude, Eq.~\eqref{eq:circular vortex amplitude details flipped}, contribute significantly to the rate. This leads to the significant differences between the cases of $\theta_p = 1~\textrm{mrad}$ (Fig.~\ref{1d_flipped_1mrad}) and $\theta_p = 0.1~\textrm{rad}$ (Fig.~\ref{1d_flipped_0.1rad}). When $\theta_p$ increases from $1~\textrm{mrad}$ to $0.1~\textrm{rad}$, not only the feature of the photon spectrum changes significantly, but also the rate increases by orders of magnitude. This change caused by transverse momenta is especially profound in the case of $\sigma=-\lambda$, since with $\sigma=\lambda$ there still exists the contribution from longitudinal momenta. Comparison with the result in Fig.~\ref{1d_xi0.1} indicates that the opening angle $\theta_p$ has a much stronger impact on the process when the electron flips its spin than preserves it. This shows the importance of the opening angle in the coupling between OAM and spin angular momentum, as well as in the manipulation of OAM and spins.

\subsection{Circularly polarized laser with $\xi\ge 1$}

Now we investigate the impact of a larger laser intensity by assuming $\xi=1,~5$, which corresponds to an intensity of about $2\times10^{18}\textrm{W}/\textrm{cm}^{2}$ and $5\times10^{19}\textrm{W}/\textrm{cm}^{2}$ respectively. We continue to work with a small opening angle $\theta_p= 1~\textrm{mrad}$. Since for such a small opening angle the rate in the electron spin-flipped case is much smaller than in the spin-preserved case, we restrict ourselves to the latter, and consider the case $\sigma=\sigma'=+1$. We also assume $\ell=\ell'=10$. 

\begin{figure}[H]
    \centering
    \includegraphics[width=0.75\linewidth]{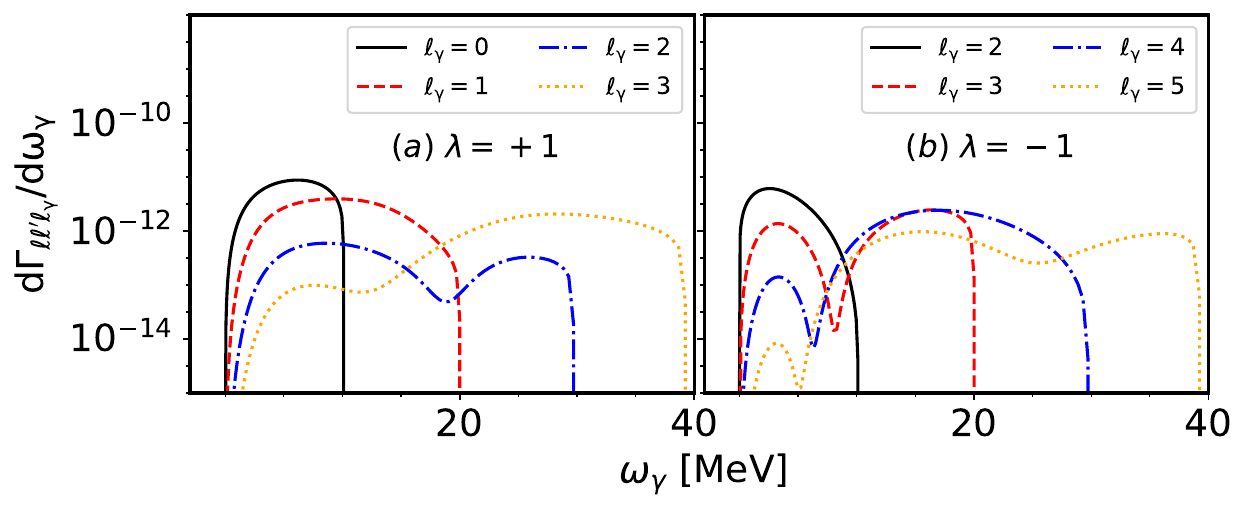}
    \caption{The distribution $d\Gamma_{\ell\ell'\ell_\gamma}/d\omega_\gamma$ of the emitted vortex $\gamma$ photon for $\lambda=+1$ ($\lambda=-1$) in panel (a) (panel (b)). In both panels $\ell=\ell'=10$, $\theta_p = 1~\textrm{mrad}$, $\sigma=\sigma'=+1$, and $\xi=1$.}
    \label{1d_xi1}
\end{figure}

\begin{figure}[H]
   \centering
    \includegraphics[width=0.9\linewidth]{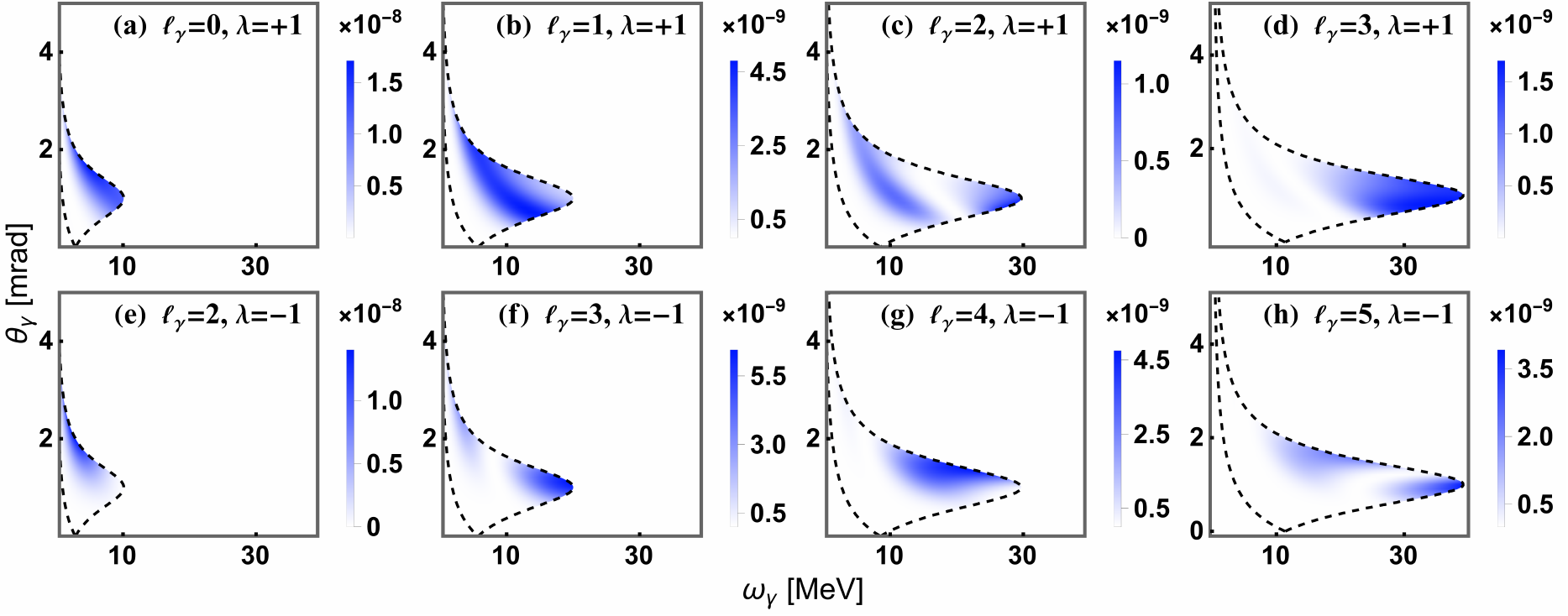}
    \caption{Same as Fig.~\ref{2d_xi0.1_1mrad} except for $\xi=1$.}
    \label{2d_xi1}
\end{figure}

The spectrum $d\Gamma_{\ell\ell'\ell_\gamma}/d\omega_\gamma$ of the emitted vortex $\gamma$ photons is presented in Fig.~\ref{1d_xi1} for the case of $\xi=1$. The rate is much larger than the one with $\xi=0.1$. Although the rate drops with the increase of $\ell_\gamma$ as before, the drop is not as significant as in the case of a smaller intensity parameter $\xi=0.1$. This has two consequences. On the one hand, it is more difficult to isolate a vortex photon state with a specific OAM. On the other, the emitted photon has a larger average OAM. We plot the two-dimensional distribution $d\Gamma_{\ell\ell'\ell_\gamma}/(d\omega_\gamma d\theta_\gamma$) in Fig.~\ref{2d_xi1} for the same set of parameters as in Fig.~\ref{1d_xi1}. As in the case of $\xi=0.1$ shown in Fig.~\ref{2d_xi0.1_1mrad}, the opening angle $\theta_\gamma$ of the emitted photon also concentrates mainly near the value of the opening angle $\theta_p$ of the initial vortex electron.

\begin{figure}[H]
    \centering
    \includegraphics[width=0.75\linewidth]{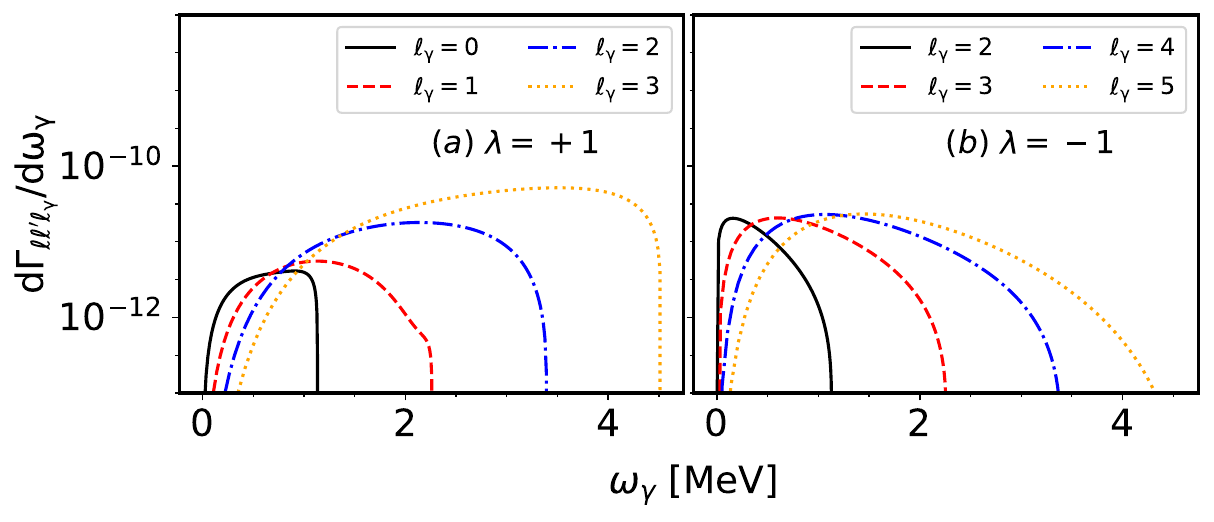}
    \caption{Same as Fig.~\ref{1d_xi1} except for $\xi=5$.}
    \label{1d_xi5}
\end{figure}

\begin{figure}[H]
   \centering
    \includegraphics[width=0.9\linewidth]{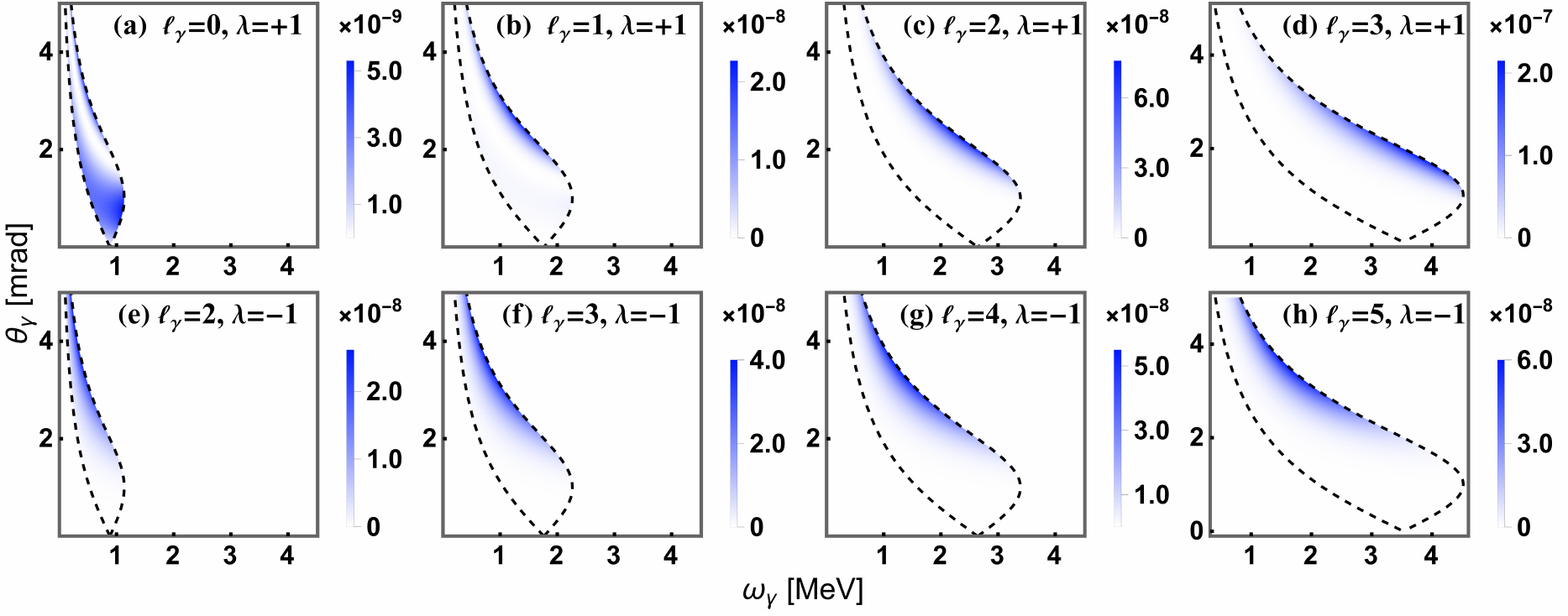}
    \caption{Same as Fig.~\ref{2d_xi1} except for $\xi=5$.}
    \label{2d_xi5}
\end{figure}

The calculation is also performed for the case $\xi=5$, with the results presented in Figs.~\ref{1d_xi5} and \ref{2d_xi5}. The rate becomes even larger, and the nonlinear effect becomes strong as demonstrated in the increase of the rate with the photon OAM $\ell_\gamma$ which demands a larger harmonic number $n$. It is also shown in Fig.~\ref{1d_xi5} and Fig.~\ref{2d_xi5} that the maximal energy of the emitted photon in the case $\xi=5$ is much smaller than the one with a smaller $\xi$. This is because a larger $\xi$ leads to a heavier dressed mass for the electron, which limits the radiated photon energy as in the plane-wave electron case. Finally, Fig.~\ref{2d_xi5} shows that the opening angle $\theta_\gamma$ of the emitted photon for most values of $\ell_\gamma$ locates in the range that is slightly larger than the opening angle $\theta_p$ of the initial vortex electron.

\subsection{Linearly polarized laser}

Since the photons of a linearly polarized laser are not in a helicity eigenstate, we cannot formulate a constraint on the initial and final particles from angular momentum conservation. In this situation the OAM $\ell_\gamma$ of the emitted photon does not have a fixed relation with the harmonic number $n$ of the laser. This means that we have to sum over $n$ for a given set of the OAMs of the initial and final particles $\ell,~\ell',~\ell_\gamma$. In numerical calculations, we calculate up to a certain value of $n$ in each case where the result has converged safely. In this circumstance we recall from Sec.~\ref{sec:varphi_integrals} that $n$ has the opposite parity to $\Delta\ell$ for the electron spin-preserved case and the same parity as $\Delta\ell$ for the spin-flipped case. Furthermore, as the Bessel function $J_\nu(\beta)$ in Eq.~\eqref{eq:linear vortex amplitude} is extremely suppressed when $\beta\ll\nu$, the summation over $\nu$ for the calculation of the amplitude can be safely truncated at a finite $\nu$.

\begin{figure}[H]
   \centering
    \includegraphics[width=0.75\linewidth]{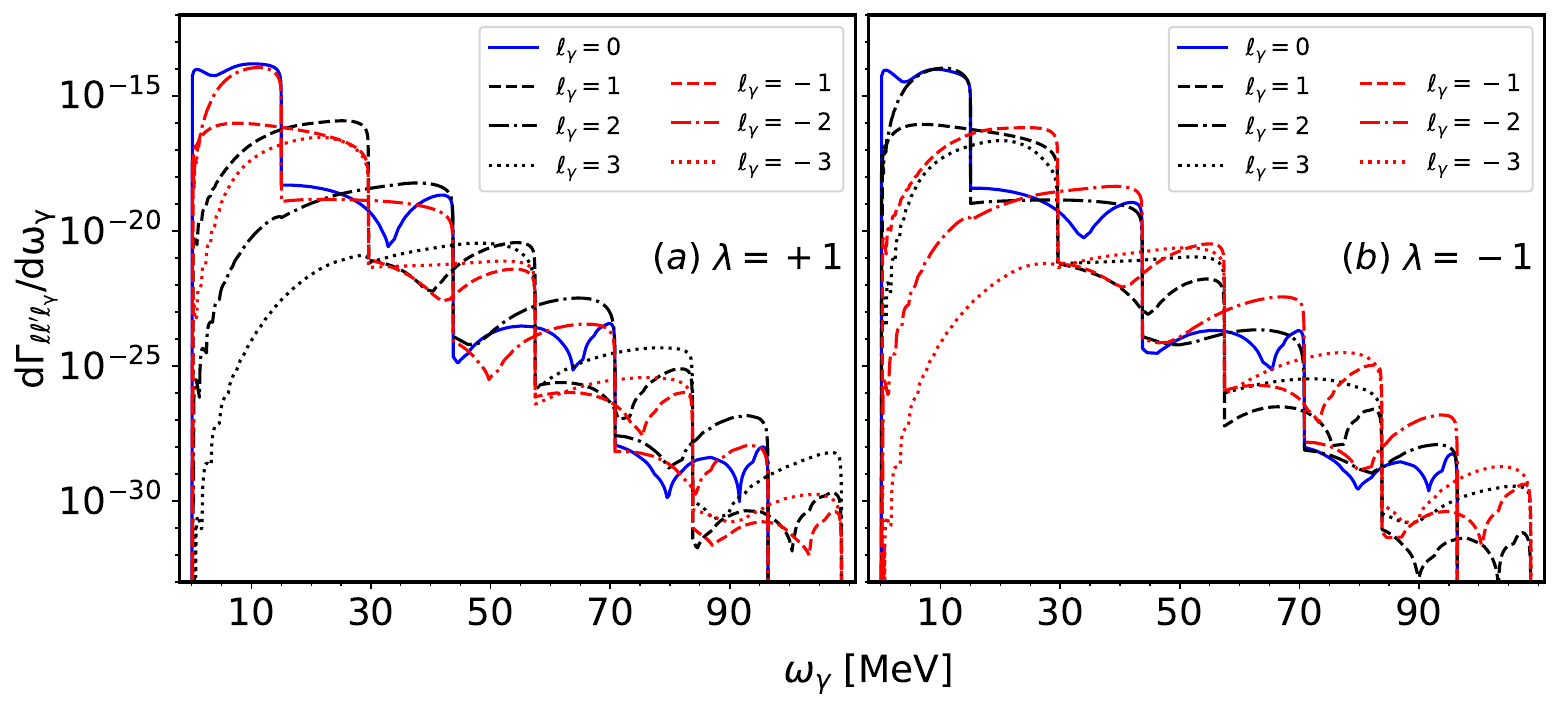}
    \caption{The distribution $d\Gamma_{\ell\ell'\ell_\gamma}/d\omega_\gamma$ of the emitted vortex $\gamma$ photon in the linearly polarized laser case and for $\lambda=+1$ ($\lambda=-1$) in panel (a) (panel (b)). In both panels $\ell=\ell'=10$, $\sigma=\sigma'=+1$, $\theta_p=1~\textrm{mrad}$, and $\xi=0.1$.}
    \label{1d_linear}
\end{figure}

Before presenting distributions we show in Table~\ref{tab:LinearRate} the total rate $\Gamma_{\ell\ell'\ell_\gamma}$ for some selected sets of $(\ell', \ell_\gamma)$ with a fixed $\ell=10$ and $\lambda=+1$. We restrict our discussion to the case when the vortex electron with a small opening angle $\theta_p = 1~\textrm{mrad}$ preserves its spin, $\sigma'=\sigma=+1$, and for a not so intense laser $\xi=0.1$. We observe that $\Gamma_{\ell\ell'\ell_\gamma}$ for small $|\ell-\ell'|$ and $|\ell_\gamma|$ is much larger than the one for a large $|\ell-\ell'|$ or $|\ell_\gamma|$. The suppression in the rate by the increase of $|\ell-\ell'|$ is similar to that in the circularly polarized laser case. The spectrum $d\Gamma_{\ell\ell'\ell_\gamma}/d\omega_\gamma$ of the emitted vortex $\gamma$ photon is presented in Fig.~\ref{1d_linear}. A multi-peak or step-function-like structure is clearly visible. This feature reveals the start-up contribution from each increasing value of $n$. On inspecting Figs.~\ref{1d_linear}(a) and (b), we observe that $d\Gamma_{\ell\ell'\ell_\gamma}/d\omega_\gamma$ with a given photon OAM $\ell_\gamma$ in the case of $\lambda=+1$ looks very similar to the one with an opposite OAM $-\ell_\gamma$ in the case of $ \lambda=-1$. As a matter of fact, the corresponding spectra coincide, as can be understood analytically through Eqs.~\eqref{linear amplitude 1}, \eqref{linear amplitude 2}, and \eqref{function G of linear polarized case}. Nonvanishing of the $G$ functions requires $\ell_\gamma$ have an opposite parity to the harmonic number $n$ when $\ell'=\ell$ as is the case shown in Fig.~\ref{1d_linear}. Furthermore, each $G$ function is a product of two Bessel functions with an identical argument but different orders which interchange when the signs of $\ell_\gamma$ and $\lambda$ are simultaneously flipped. Finally, within the energy range of each step no single $\ell_\gamma$ dominates the spectrum. This is a reflection of the fact that there is no manifest angular momentum conservation in a linearly polarized laser, and implies that it would be difficult to isolate a single vortex photon with a given OAM in such a case.

\begin{table} [h]
\centering
\renewcommand\arraystretch{1.6}
\begin{tabular} {lcccccc}
  \hline\hline
   $(\ell', \ell_\gamma)$ & $(10,0)$ & $(10, 1)$ & $(10, 10)$ & $(10, -10)$ & $(0, 1)$ & $(20, 1)$ \\
  \hline
   $\Gamma_{\ell\ell'\ell_\gamma}$ [MeV]~~~& $1.7 \times 10^{-13}$ & $1.6 \times 10^{-15}$ & $5.3 \times 10^{-34}$ & $8.6 \times 10^{-31}$ & $4.0 \times 10^{-27}$ & $5.1 \times 10^{-35}$\\
  \hline\hline
\end{tabular}
\renewcommand\arraystretch{1}
\caption{The total rate $\Gamma_{\ell\ell'\ell_\gamma}$ in a linearly polarized laser for some value of $(\ell',\ell_\gamma)$. Other parameters are, $\ell=10$, $\sigma=\sigma'=+1$, $\lambda=+1$, $\theta_p=1~\textrm{mrad}$, and $\xi=0.1$.}
\label{tab:LinearRate}
\end{table}

\section{Conclusion}
\label{sec:conclusion}

We have studied the process of all-vortex nonlinear Compton scattering in an intense and polarized laser field, in which the initial and final electrons and the emitted $\gamma$ photon are all in vortex states. We have developed a formalism for the process, which allows us to study the exchanges of the OAM and spin angular momentum among the electron, $\gamma$ photon, and laser. Both circularly and linearly polarized lasers have been investigated.

Our numerical results have shown that, with a less intense laser intensity ($\xi=0.1$) and a relativistic initial electron, the spectrum $d\Gamma_{\ell\ell'\ell_\gamma}/d\omega_\gamma$ of the emitted vortex $\gamma$ photon has a strong dependence on the OAM $\ell_\gamma$ of the emitted photon and can change by orders of magnitude with different $\ell_\gamma$. Together with the strong $\ell_\gamma$ dependence of the range of the energy of the emitted photon, this could offer a possible scenario to separate emitted vortex $\gamma$ photons with different OAMs. A multi-peak structure in the spectrum was found, which depends strongly on the OAMs of the involved particles and the spins of the electrons, as well as the helicity of the $\gamma$ photon. This could offer a possible way to identify the OAM and spin of the particle. The suppression on the rate and $d\Gamma_{\ell\ell'\ell_\gamma}/d\omega_\gamma$ from the difference between the OAMs of the initial and final vortex electrons has been revealed also. The strong impact of the opening angle of the initial vortex electron has been demonstrated, especially for the electron spin-flipped case. For a high laser intensity ($\xi=1$ or $\xi=5$), our numerical calculations have shown that the nonlinear effect is strong. The rate generally increases with the laser intensity. Different from the low laser intensity case ($\xi=0.1$), the order of magnitude of the spectrum keeps similar (at $\xi=1$) or even increases (at $\xi=5$) when $\ell_\gamma$ increases. This implies that a higher laser intensity does not help to isolate an emitted vortex photon with a given definite OAM although it induces a larger average OAM.

\section*{Acknowledgements}
This work was supported in part by the Grants 
No.\,NSFC-12035008 and No.\,NSFC-12475122, 
by the Guangdong Major Project of Basic and Applied Basic Research No.\,2020B0301030008, and by the Fundamental Research Funds for the Central Universities (Grant No. 010-63253121).

\appendix

\section{Calculation of the current for a Bessel-Volkov solution}
\label{sec:appendix_A}

The Bessel-Volkov solution in Eq.~\eqref{eq:initial_e} reads in light-front coordinates: 
\begin{eqnarray}
\psi_{p_+\kappa\ell\sigma}^{(a)}(x)
&=&N_p^\B e^{-i(x_+p_-+x_-p_+)/2}e^{i\phi_a^p}i^{-\ell}\sqrt{\frac{\kappa}{2\pi}}
\left(1+\frac{\slashed{\Omega}\slashed{a}}{2p\cdot\Omega}\right)W,
\label{eq:psi_lightfront}
\end{eqnarray}
where \cite{Karlovets:2012eu}
\begin{subequations}
\begin{eqnarray}
&&\phi_a^p=\int^\tau d\tau'\frac{a^2(\tau')}{2p\cdot\Omega},
\\
&&W=\int\frac{d\phi_p}{2\pi}e^{i\ell\phi_p}e^{i\Xi}u(p,\sigma),
\\
&&\Xi=\pvec_\perp\cdot\bfcalR_\perp
=\kappa\calR_\perp\cos(\phi_p-\phi_\calR),~~~
\\
&&\bfcalR_\perp=\xvec_\perp+\frac{1}{p\cdot\Omega}\int^\tau d\tau'~\avec_\perp(\tau'),
\end{eqnarray}
\end{subequations}
with $\phi_\calR$ being the azimuthal angle of $\bfcalR_\perp$. Including the $e^{\pm i\phi_p}$ phases from $u(p,\sigma)$, the $\phi_p$ integral is finished by shifting $\phi_p\to\phi_p+\phi_\calR$ and in terms of the Bessel function, 
\begin{eqnarray}
\int_0^{2\pi}\frac{d\phi}{2\pi}~e^{i(n\phi+x\cos\phi)}=i^nJ_n(x).
\end{eqnarray}
The result is \cite{Bliokh:2011fi}
\begin{eqnarray}
W&=&\left(\begin{array}{c}
\epsilon_+w_{\sigma} 
\\
\epsilon_-\sigma\cos\theta_{p}w_{\sigma}
\end{array}\right)e^{i\ell\phi_\calR}J_{\ell}
+\left(\begin{array}{c}
0 
\\
w_{-\sigma}\end{array}\right)
i^\sigma e^{i(\ell+\sigma)\phi_\calR}J_{\ell+\sigma}\epsilon_-\sin\theta_p,
\end{eqnarray}
where all Bessel functions share the argument $\kappa\calR_\perp$, $\epsilon_\pm=\sqrt{E\pm M}$, and $\theta_p$ is the (half) cone angle of the Bessel vortex. Note that $W$ has a definite value $\ell+\sigma/2$ of the total angular momentum in the $z$ axis. To form the plus component of the current $j_+
=\bar\psi_{p_+\kappa\ell\sigma}^{(a)}\gamma_+\psi_{p_+\kappa\ell\sigma}^{(a)}$, we recall that the laser propagates in the $(-z)$ direction ($\Omega_+=\Omega_\perp=0$) so that $\slashed{\Omega}=\Omega_-\gamma_+/2,~\slashed{\Omega}\gamma_+=\gamma_+\slashed{\Omega}=0$, thus 
\begin{eqnarray}
j_+&=&(N_p^\B)^2\frac{\kappa}{2\pi}\overline{W}\gamma_+W
\nonumber
\\
&=&(N_p^\B)^2\frac{\kappa}{2\pi}\Big[
(\epsilon_++\epsilon_-\cos\theta_p)^2\big(J_\ell\big)^2
+\epsilon_-^2\sin^2\theta_p\big(J_{\ell+\sigma}\big)^2\Big].
\end{eqnarray}
Requiring the normalization condition 
\begin{eqnarray}
\frac{1}{2}\int dx_-d^2\xvec_\perp~j_+=1, 
\label{eq:current_nor}
\end{eqnarray}
and noting that $\bfcalR_\perp$ differs from $\xvec_\perp$ by a term that does not depend on $\xvec_\perp$, the condition becomes 
\begin{eqnarray}
1=\frac{1}{2}(N_p^\B)^2\frac{\kappa}{2\pi}\int dx_-2\pi \calR_\perp d\calR_\perp\Big[
(\epsilon_++\epsilon_-\cos\theta_p)^2\big(J_\ell\big)^2
+\epsilon_-^2\sin^2\theta_p\big(J_{\ell+\sigma}\big)^2
\Big].  
\end{eqnarray}
Regularizing the minus coordinate by a large $L_-$ and the radius of the transverse plane by a large $R$, and using 
\begin{eqnarray}
\int_0^R rdr\big(J_\ell(\kappa r)\big)^2=\frac{R}{\pi\kappa},~~~R\to\infty,
\end{eqnarray}
we finally obtain 
\begin{eqnarray}
N_p^\B=\frac{1}{\sqrt{2p_+}}\sqrt{\frac{2\pi}{RL_-}},
\end{eqnarray}
which is the same as that in free space, Eq,~\eqref{eq:norm_free}.

\section{Planar integrals}
\label{sec:appendix_B}

The $\delta$ function in planar polar coordinates can be determined from its Descartes form as follows: 
\begin{subequations}
\begin{eqnarray}
&&\iint d^2\xvec~\delta^2(\xvec-\xvec_0)f(\xvec)
=f(\xvec_0),
\\
&&\iint \rho d\rho d\varphi~2\delta(\rho^2-\rho_0^2)\delta(\varphi-\varphi_0)f(\rho,\varphi)
\nonumber
\\
&&=\iint d\rho^2 d\varphi~\delta(\rho^2-\rho_0^2)\delta(\varphi-\varphi_0)f(\rho,\varphi)
=f(\rho_0,\varphi_0),
\\
\Longrightarrow&&\delta^2(\xvec-\xvec_0)
=2\delta(\rho^2-\rho_0^2)\delta(\varphi-\varphi_0)
\nonumber
\\
&&=\rho_0^{-1}\delta(\rho-\rho_0)\delta(\varphi-\varphi_0)
\textrm{ for }\rho_0>0.
\end{eqnarray}
\end{subequations}
Consider the planar integral \cite{Ivanov:2011kk}: 
\begin{eqnarray}
J&=&\iint d\varphi_1d\varphi_2~\delta^2(\kvec_1+\kvec_2-\kvec)f(\varphi_1,\varphi_2).
\end{eqnarray}
The $\delta$ function requires the three vectors to form a triangle. Given their magnitudes, there are two solutions as can be seen geometrically: Drawing one circle of radius $|\kvec_1|=k_1$ at one end of $\kvec$ and another circle of radius $|\kvec_2|=k_2$ at the other end, the two circles intersect at two points which are the solutions: 
\begin{subequations}
\begin{eqnarray}
&&\angle(\kvec,\kvec_{1,2})=\delta_{1,2}
=\arccos\frac{k^2+k_{1,2}^2-k_{2,1}^2}{2kk_{1,2}},
\\
&&
2\kvec\cdot\kvec_{1,2}=k^2+k_{1,2}^2-k_{2,1}^2,
\\
&&
\varphi_k-\varphi_1=\pm\delta_1,~~~\varphi_k-\varphi_2=\mp\delta_2.
\end{eqnarray}
\end{subequations}
Then, we have 
\begin{eqnarray}
J&=&\iint d\varphi_1d\varphi_2~2\delta\big(\kvec_1^2-(\kvec-\kvec_2)^2\big)\delta(\varphi_1-\varphi_{\kvec-\kvec_2})f(\varphi_1,\varphi_2)
\nonumber
\\
&=&2\int d\varphi_2~\delta\big(k_1^2-k^2-k_2^2+2kk_2\cos(\varphi_2-\varphi_k)\big)f(\varphi_{\kvec-\kvec_2},\varphi_2)
\nonumber
\\
&=&\frac{1}{2\Delta}\big[f(\varphi_k-\delta_1,\varphi_k+\delta_2)+f(\varphi_k+\delta_1,\varphi_k-\delta_2)\big],
\end{eqnarray}
where $\Delta$ is the area of the triangle: 
\begin{eqnarray}
\Delta=\frac{1}{2}kk_2\sin\delta_2
=\frac{1}{4}\sqrt{-\lambda(k_1^2,k_2^2,k^2)}.
\end{eqnarray}

\section{Basic angular integral in a linearly polarized laser}
\label{sec:appendix_C}

The basic azimuthal integral to be evaluated in a linearly polarized case is, 
\begin{eqnarray}
G_\mu^m(\eta)
&=&\int_0^{2\pi} d\varphi~e^{im\varphi}J_\mu\big(\gamma^\rmL\big),
\end{eqnarray}
where $\gamma^\rmL=\sqrt{2}\gamma^\rmC\cos(\varphi-\eta\varphi_0)$, $m,~\mu$ are both integers, $\eta=\pm 1$, and only the parameters relevant to the calculation of the integral are indicated in $G$. Shifting $\varphi\to\varphi+\eta\varphi_0$ gives 
\begin{subequations}
\begin{eqnarray}
G_\mu^m(\eta)
&=&e^{im\eta\varphi_0}H_\mu^m,
\\
H_\mu^m&=&\int_b^{b+2\pi} d\varphi~e^{im\varphi}
J_\mu\big(2a\cos\varphi\big),
\end{eqnarray}
\end{subequations}
where $a=\gamma^\rmC/\sqrt{2}$ and $b=-\eta\varphi_0$. The integrand of $H_\mu^m$ is a periodic function of $\varphi$ in the interval and thus $H_\mu^m$ actually does not depend on the choice of $b$. To apply formulae (6.681)-8 and -9 in Ref.\cite{Gradshteyn:2007}, we first focus on the case with integers $\mu\ge 0$ and $m$ arbitrary, and will come back to the case of $\mu<0$ by using $J_{-\mu}(z)=(-1)^\mu J_\mu(z)$. Choosing $b=\pi/2$, shifting $\varphi\to\varphi+\pi/2$, and then using $J_\mu(-z)=(-1)^\mu J_\mu(z)$, we have 
\begin{eqnarray}
H_\mu^m&=&i^m(-1)^\mu\int_0^{2\pi}d\varphi~e^{im\varphi}
J_\mu\big(2a\sin\varphi\big).
\end{eqnarray}
We cut the interval into two halves, and for the upper half we shift $\varphi\to\varphi+\pi$, so that 
\begin{eqnarray}
H_\mu^m&=&
i^m[(-1)^\mu+(-1)^m]\int_0^\pi d\varphi~e^{im\varphi}J_\mu\big(2a\sin\varphi\big)
\nonumber
\\
&=&i^m\big[(-1)^\mu+(-1)^m\big]i^m\pi 
J_{(\mu+m)/2}(a)J_{(\mu-m)/2}(a)
\nonumber
\\
&=&\big[1+(-1)^{\mu+m}\big]\pi J_{(\mu+m)/2}(a)J_{(\mu-m)/2}(a).
\end{eqnarray}
In summary, for $\mu\ge 0$, 
\begin{eqnarray}
\label{function G of linear polarized case}
G_\mu^m(\eta)
=\big[1+(-1)^{\mu+m}\big]e^{im\eta\varphi_0}
\pi J_{(\mu+m)/2}(\gamma^\rmC/\sqrt{2})
J_{(\mu-m)/2}(\gamma^\rmC/\sqrt{2}),
\end{eqnarray}
which requires integers $\mu,~m$ to have the same parity to be nonvanishing. For $\mu<0$, we use $J_{-\mu}(z)=(-1)^\mu J_\mu(z)$ to obtain 
\begin{eqnarray}
G_\mu^m(\eta)
&=&e^{im\eta\varphi_0}(-1)^\mu\int_0^{2\pi} d\varphi~e^{im\varphi}
J_{-\mu}\big(\sqrt{2}\gamma^\rmC\cos\varphi\big)
\nonumber
\\
&=&e^{im\eta\varphi_0}(-1)^\mu
\big[1+(-1)^{-\mu+m}\big]\pi J_{(-\mu+m)/2}(\gamma^\rmC/\sqrt{2})
J_{(-\mu-m)/2}(\gamma^\rmC/\sqrt{2})
\nonumber
\\
&=&e^{im\eta\varphi_0}
\big[1+(-1)^{\mu+m}\big]\pi J_{(\mu+m)/2}(\gamma^\rmC/\sqrt{2})
J_{(\mu-m)/2}(\gamma^\rmC/\sqrt{2}),
\end{eqnarray}
i.e., the result generally applies for any integers $\mu$ and $m$. 

\section{Integral over $\kappa$}
\label{sec:appendix_D}

We have transformed in section \ref{sec:rate_gauss} the original integral over $\kappa$ to the following form: 
\begin{eqnarray}
I&=&\int_{x_1}^{x_2}dx\frac{g(x)}{\sqrt{(x_2-x)(x-x_1)}},
\end{eqnarray}
where $g(x)$ is a smooth function. To isolate the apparent singularities at the endpoints, we make subtractions: 
\begin{eqnarray}
I&=&\int_{x_1}^{x_2}dx\left(\frac{1}{\sqrt{x_2-x}}-\frac{1}{\sqrt{x_2-x_1}}\right)
\left(\frac{1}{\sqrt{x-x_1}}-\frac{1}{\sqrt{x_2-x_1}}\right)g(x)
\nonumber
\\
&&+\frac{1}{\sqrt{x_2-x_1}}\int_{x_1}^{x_2}dx
\left(\frac{1}{\sqrt{x-x_1}}+\frac{1}{\sqrt{x_2-x}}\right)g(x)
\nonumber
\\
&&
-\frac{1}{x_2-x_1}\int_{x_1}^{x_2}dx~g(x).
\end{eqnarray}
The integrand in the first integral vanishes at the endpoints and is thus well behaved. Each integrand in the second integral is singular only at one endpoint and is easy to cope with. It can be further softened by using integration by parts: 
\begin{subequations}
\begin{eqnarray}
&&\int_{x_1}^{x_2}dx\frac{g(x)}{\sqrt{x-x_1}}
=2g(x_2)\sqrt{x_2-x_1}-2\int dx~g'(x)\sqrt{x-x_1},
\\
&&\int_{x_1}^{x_2}dx\frac{g(x)}{\sqrt{x_2-x}}
=2g(x_1)\sqrt{x_2-x_1}+2\int dx~g'(x)\sqrt{x_2-x}.
\end{eqnarray}
\end{subequations}
To make the first integral numerically amenable, we combine the square roots: 
\begin{eqnarray}
&&\left(\frac{1}{\sqrt{x_2-x}}-\frac{1}{\sqrt{x_2-x_1}}\right)
\left(\frac{1}{\sqrt{x-x_1}}-\frac{1}{\sqrt{x_2-x_1}}\right)
\nonumber
\\
&=&\frac{1}{x_2-x_1}
\frac{\sqrt{(x_2-x)(x-x_1)}}{\left(\sqrt{x_2-x_1}+\sqrt{x_2-x}\right)
\left(\sqrt{x_2-x_1}+\sqrt{x-x_1}\right)}.
\end{eqnarray}
In summary, the final form is, 
\begin{eqnarray}
I&=&-\frac{1}{x_2-x_1}\int_{x_1}^{x_2}dx
\frac{(x_2-x_1)+\sqrt{x_2-x_1}\big(\sqrt{x-x_1}+\sqrt{x_2-x}\big)}{\big(\sqrt{x_2-x_1}+\sqrt{x_2-x}\big)
\big(\sqrt{x_2-x_1}+\sqrt{x-x_1}\big)}g(x)
\nonumber
\\
&&+\frac{2}{\sqrt{x_2-x_1}}\int dx~g'(x)(\sqrt{x_2-x}-\sqrt{x-x_1})
+2\big[g(x_1)+g(x_2)\big].
\label{eq:I_result}
\end{eqnarray}

\bibliography{paper250406ref}

\providecommand{\href}[2]{#2}\begingroup\raggedright\begin{thebibliography}{100}

\bibitem{Bliokh:2015doa}
K.~Y. Bliokh and F.~Nori, ``{\it {Transverse and longitudinal angular momenta of light}},'' \href{http://dx.doi.org/10.1016/j.physrep.2015.06.003}{{\em Phys. Rept.} {\bfseries 592} (2015) 1--38}, [\href{http://arxiv.org/abs/1504.03113}{{\ttfamily arXiv:1504.03113}} [physics.optics]].

\bibitem{Knyazev:2018}
B.~A. Knyazev and V.~G. Serbo, ``{\it Beams of photons with nonzero projections of orbital angular momenta: new results},'' \href{http://dx.doi.org/10.3367/UFNe.2018.02.038306}{{\em Physics-Uspekhi} {\bfseries 61} (2018) 449}.

\bibitem{Bliokh:2017uvr}
K.~Y. Bliokh {\em et~al.}, ``{\it Theory and applications of free-electron vortex states},'' \href{http://dx.doi.org/10.1016/j.physrep.2017.05.006}{{\em Phys. Rept.} {\bfseries 690} (2017) 1--70}, [\href{http://arxiv.org/abs/1703.06879}{{\ttfamily arXiv:1703.06879}} [quant-ph]].

\bibitem{Yuan:2017}
S.~M. Lloyd, M.~Babiker, G.~Thirunavukkarasu, and J.~Yuan, ``{\it Electron vortices: Beams with orbital angular momentum},'' \href{http://dx.doi.org/10.1103/RevModPhys.89.035004}{{\em Rev. Mod. Phys.} {\bfseries 89} (2017) 035004}.

\bibitem{Ivanov:2022jzh}
I.~P. Ivanov, ``{\it {Promises and challenges of high-energy vortex states collisions}},'' \href{http://dx.doi.org/10.1016/j.ppnp.2022.103987}{{\em Prog. Part. Nucl. Phys.} {\bfseries 127} (2022) 103987}, [\href{http://arxiv.org/abs/2205.00412}{{\ttfamily arXiv:2205.00412}} [hep-ph]].

\bibitem{Yao:11}
A.~M. Yao and M.~J. Padgett, ``{\it Orbital angular momentum: origins, behavior and applications},'' \href{http://dx.doi.org/10.1364/AOP.3.000161}{{\em Adv. Opt. Photon.} {\bfseries 3} (2011) 161}.

\bibitem{shen:2019}
Y.~Shen, X.~Wang, Z.~Xie, C.~Min, X.~Fu, Q.~Liu, M.~Gong, and X.~Yuan, ``{\it Optical vortices 30 years on: Oam manipulation from topological charge to multiple singularities},'' {\em Light: Science \& Applications} {\bfseries 8} (2019) 90.

\bibitem{Forbes:2024}
A.~Forbes, L.~Mkhumbuza, and L.~Feng, ``{\it Orbital angular momentum lasers},'' {\em Nature Reviews Physics} {\bfseries 6} (2024) 352.

\bibitem{Allen:1992zz}
L.~Allen, M.~W. Beijersbergen, R.~J.~C. Spreeuw, and J.~P. Woerdman, ``{\it {Orbital angular momentum of light and the transformation of Laguerre-Gaussian laser modes}},'' \href{http://dx.doi.org/10.1103/PhysRevA.45.8185}{{\em Phys. Rev. A} {\bfseries 45} (1992) 8185--8189}.

\bibitem{BEIJERSBERGEN:1994}
M.~Beijersbergen, R.~Coerwinkel, M.~Kristensen, and J.~Woerdman, ``{\it Helical-wavefront laser beams produced with a spiral phaseplate},'' \href{http://dx.doi.org/https://doi.org/10.1016/0030-4018(94)90638-6}{{\em Optics Communications} {\bfseries \textbf{112}} (1994) 321}.

\bibitem{Oemrawsingh:2004}
S.~S.~R. Oemrawsingh, J.~A.~W. van Houwelingen, E.~R. Eliel, J.~P. Woerdman, E.~J.~K. Verstegen, J.~G. Kloosterboer, and G.~W. 't~Hooft, ``{\it Production and characterization of spiral phase plates for optical wavelengths},'' \href{http://dx.doi.org/10.1364/AO.43.000688}{{\em Appl. Opt.} {\bfseries \textbf{43}} (2004) 688}.

\bibitem{Ruffato:2014}
G.~Ruffato, M.~Massari, and F.~Romanato, ``{\it Generation of high-order laguerre gaussian modes by means of spiral phase plates},'' \href{http://dx.doi.org/10.1364/OL.39.005094}{{\em Opt. Lett.} {\bfseries \textbf{39}} (2014) 5094}.

\bibitem{Longman:2020}
A.~Longman, C.~Salgado, G.~Zeraouli, J.~I.~A. {n}aniz, J.~A. P\'{e}rez-Hern\'{a}ndez, M.~K. Eltahlawy, L.~Volpe, and R.~Fedosejevs, ``{\it Off-axis spiral phase mirrors for generating high-intensity optical vortices},'' \href{http://dx.doi.org/10.1364/OL.387363}{{\em Opt. Lett.} {\bfseries \textbf{45}} (2020) 2187}.

\bibitem{Heckenberg:1992}
N.~R. Heckenberg, R.~McDuff, C.~P. Smith, and A.~G. White, ``{\it Generation of optical phase singularities by computer-generated holograms},'' \href{http://dx.doi.org/10.1364/OL.17.000221}{{\em Opt. Lett.} {\bfseries \textbf{17}} (1992) 221}.

\bibitem{Peele:2002}
A.~G. Peele, P.~J. McMahon, D.~Paterson, C.~Q. Tran, A.~P. Mancuso, K.~A. Nugent, J.~P. Hayes, E.~Harvey, B.~Lai, and I.~McNulty, ``{\it Observation of an x-ray vortex},'' \href{http://dx.doi.org/10.1364/OL.27.001752}{{\em Opt. Lett.} {\bfseries \textbf{27}} (2002) 1752}.

\bibitem{Terhalle:2011}
B.~Terhalle, A.~Langner, B.~P\"{a}iv\"{a}nranta, V.~A. Guzenko, C.~David, and Y.~Ekinci, ``{\it Generation of extreme ultraviolet vortex beams using computer generated holograms},'' \href{http://dx.doi.org/10.1364/OL.36.004143}{{\em Opt. Lett.} {\bfseries \textbf{36}} (2011) 4143}.

\bibitem{Gariepy:2014}
G.~Gariepy, J.~Leach, K.~T. Kim, T.~J. Hammond, E.~Frumker, R.~W. Boyd, and P.~B. Corkum, ``{\it Creating high-harmonic beams with controlled orbital angular momentum},'' \href{http://dx.doi.org/10.1103/PhysRevLett.113.153901}{{\em Phys. Rev. Lett.} {\bfseries 113} (2014) 153901}.

\bibitem{Hemsing:2013}
E.~Hemsing, A.~Knyazik, M.~Dunning, D.~Xiang, A.~Marinelli, C.~Hast, and J.~B. Rosenzweig, ``{\it Coherent optical vortices from relativistic electron beams},'' \href{http://dx.doi.org/10.1038/nphys2712}{{\em Nature Physics} {\bfseries 9} (2013) 549}.

\bibitem{Bahrdt:2013eoa}
J.~Bahrdt, K.~Holldack, P.~Kuske, R.~M\"uller, M.~Scheer, and P.~Schmid, ``{\it {First Observation of Photons Carrying Orbital Angular Momentum in Undulator Radiation}},'' \href{http://dx.doi.org/10.1103/PhysRevLett.111.034801}{{\em Phys. Rev. Lett.} {\bfseries 111} (2013) 034801}.

\bibitem{Gauthier:2016}
D.~C. Gauthier {\em et~al.}, ``{\it Tunable orbital angular momentum in high-harmonic generation},'' \href{http://dx.doi.org/10.1038/ncomms14971}{{\em Nature Communications} {\bfseries 8} (2016) 14971}.

\bibitem{Lee:2019}
J.~C.~T. Lee, S.~J. Alexander, S.~D. Kevan, S.~Roy, and B.~J. McMorran, ``{\it Laguerre–gauss and hermite–gauss soft x-ray states generated using diffractive optics},'' \href{http://dx.doi.org/10.1038/s41566-018-0328-8}{{\em Nature Photonics} {\bfseries 13} (2019) 205}.

\bibitem{Uchida:2010hbm}
M.~Uchida and A.~Tonomura, ``{\it {Generation of electron beams carrying orbital angular momentum}},'' \href{http://dx.doi.org/10.1038/nature08904}{{\em Nature} {\bfseries 464} no.~7289, (2010) 737--739}.

\bibitem{Verbeeck:2010ezk}
J.~Verbeeck, H.~Tian, and P.~Schattschneider, ``{\it {Production and application of electron vortex beams}},'' \href{http://dx.doi.org/10.1038/nature09366}{{\em Nature} {\bfseries 467} no.~7313, (2010) 301--304}.

\bibitem{McMorran:2011bql}
B.~J. McMorran, A.~Agrawal, I.~M. Anderson, A.~A. Herzing, H.~J. Lezec, J.~J. McClelland, and J.~Unguris, ``{\it {Electron Vortex Beams with High Quanta of Orbital Angular Momentum}},'' \href{http://dx.doi.org/10.1126/science.1198804}{{\em Science} {\bfseries 331} no.~6014, (2011) 1198804}.

\bibitem{Mafakheri:2017}
E.~Mafakheri, A.~H. Tavabi, P.-H. Lu, R.~Balboni, F.~Venturi, C.~Menozzi, G.~C. Gazzadi, S.~Frabboni, A.~Sit, R.~E. Dunin-Borkowski, E.~Karimi, and V.~Grillo, ``{\it Realization of electron vortices with large orbital angular momentum using miniature holograms fabricated by electron beam lithography},'' \href{http://dx.doi.org/10.1063/1.4977879}{{\em Applied Physics Letters} {\bfseries 110} (2017) 093113}.

\bibitem{Tavabi:2022}
A.~H. Tavabi, P.~Rosi, A.~Roncaglia, E.~Rotunno, M.~Beleggia, P.-H. Lu, L.~Belsito, G.~Pozzi, S.~Frabboni, P.~Tiemeijer, R.~E. Dunin-Borkowski, and V.~Grillo, ``{\it Generation of electron vortex beams with over 1000 orbital angular momentum quanta using a tunable electrostatic spiral phase plate},'' \href{http://dx.doi.org/10.1063/5.0093411}{{\em Applied Physics Letters} {\bfseries 121} (2022) 073506}.

\bibitem{Vanacore:2019}
G.~M. Vanacore, G.~Berruto, I.~Madan, E.~Pomarico, P.~Biagioni, R.~Lamb, D.~McGrouther, O.~Reinhardt, I.~Kaminer, B.~Barwick, {\em et~al.}, ``{\it Ultrafast generation and control of an electron vortex beam via chiral plasmonic near fields},'' \href{http://dx.doi.org/10.1038/s41563-019-0336-1}{{\em Nat. Mater.} {\bfseries 18} (2019) 573}, [\href{http://arxiv.org/abs/1806.00366}{{\ttfamily arXiv:1806.00366}} [quant-ph]].

\bibitem{Clark:2015rcq}
C.~W. Clark, R.~Barankov, M.~G. Huber, M.~Arif, D.~G. Cory, and D.~A. Pushin, ``{\it {Controlling neutron orbital angular momentum}},'' \href{http://dx.doi.org/10.1038/nature15265}{{\em Nature} {\bfseries 525} no.~7570, (2015) 504--506}.

\bibitem{Sarenac:2019}
D.~Sarenac, C.~Kapahi, W.~Chen, C.~W. Clark, D.~G. Cory, M.~G. Huber, I.~Taminiau, K.~Zhernenkov, and D.~A. Pushin, ``{\it Generation and detection of spin-orbit coupled neutron beams},'' \href{http://dx.doi.org/10.1073/pnas.1906861116}{{\em Proceedings of the National Academy of Sciences} {\bfseries 116} no.~41, (2019) 20328--20332}.

\bibitem{Sarenac:2022}
D.~Sarenac, M.~E. Henderson, H.~Ekinci, C.~W. Clark, D.~G. Cory, L.~DeBeer-Schmitt, M.~G. Huber, C.~Kapahi, and D.~A. Pushin, ``{\it Experimental realization of neutron helical waves},'' \href{http://dx.doi.org/10.1126/sciadv.add2002}{{\em Science Advances} {\bfseries 8} no.~46, (2022) eadd2002}.

\bibitem{luski:2021}
A.~Luski, Y.~Segev, R.~David, O.~Bitton, H.~Nadler, A.~R. Barnea, A.~Gorlach, O.~Cheshnovsky, I.~Kaminer, and E.~Narevicius, ``{\it Vortex beams of atoms and molecules},'' \href{http://dx.doi.org/10.1126/science.abj2451}{{\em Science} {\bfseries 373} no.~6559, (2021) 1105--1109}.

\bibitem{mair:2001}
A.~Mair, A.~Vaziri, G.~Weihs, and A.~Zeilinger, ``{\it Entanglement of the orbital angular momentum states of photons},'' \href{http://dx.doi.org/10.1038/35085529}{{\em Nature} {\bfseries 412} no.~6844, (2001) 313--316}.

\bibitem{Wang:2015vrl}
X.-L. Wang, X.-D. Cai, Z.-E. Su, M.-C. Chen, D.~Wu, L.~Li, N.-L. Liu, C.-Y. Lu, and J.-W. Pan, ``{\it {Quantum teleportation of multiple degrees of freedom of a single photon}},'' \href{http://dx.doi.org/10.1038/nature14246}{{\em Nature} {\bfseries 518} no.~7540, (2015) 516--519}.

\bibitem{He:1995}
H.~He, M.~E.~J. Friese, N.~R. Heckenberg, and H.~Rubinsztein-Dunlop, ``{\it Direct observation of transfer of angular momentum to absorptive particles from a laser beam with a phase singularity},'' \href{http://dx.doi.org/10.1103/PhysRevLett.75.826}{{\em Phys. Rev. Lett.} {\bfseries 75} (1995) 826--829}.

\bibitem{Simpson:1997}
N.~B. Simpson, K.~Dholakia, L.~Allen, and M.~J. Padgett, ``{\it Mechanical equivalence of spin and orbital angular momentum of light: an optical spanner},'' \href{http://dx.doi.org/10.1364/OL.22.000052}{{\em Opt. Lett.} {\bfseries 22} no.~1, (1997) 52--54}.

\bibitem{Grier:2003}
D.~G. {Grier}, ``{\it {A revolution in optical manipulation}},'' \href{http://dx.doi.org/10.1038/nature01935}{{\em nature} {\bfseries 424} no.~6950, (2003) 810--816}.

\bibitem{Larocque:2018}
H.~Larocque, I.~Kaminer, V.~Grillo, R.~W. Boyd, and E.~Karimi, ``{\it Twisting neutrons may reveal their internal structure},'' \href{http://dx.doi.org/10.1038/nphys4322x}{{\em Nature Physics} {\bfseries 14} no.~1, (2018) 1--2}.

\bibitem{kaminer:2015}
I.~Kaminer, J.~Nemirovsky, M.~Rechtsman, R.~Bekenstein, and M.~Segev, ``{\it Self-accelerating dirac particles and prolonging the lifetime of relativistic fermions},'' \href{http://dx.doi.org/10.1038/nphys3196}{{\em Nature Physics} {\bfseries 11} no.~3, (2015) 261--267}.

\bibitem{madan:2020}
I.~Madan, G.~M. Vanacore, S.~Gargiulo, T.~LaGrange, and F.~Carbone, ``{\it The quantum future of microscopy: Wave function engineering of electrons, ions, and nuclei},'' \href{http://dx.doi.org/10.1063/1.5143008}{{\em Applied Physics Letters} {\bfseries 116} (2020) 230502}.

\bibitem{Picón:2010}
A.~Picón, A.~Benseny, J.~Mompart, J.~R. Vázquez~de Aldana, L.~Plaja, G.~F. Calvo, and L.~Roso, ``{\it Transferring orbital and spin angular momenta of light to atoms},'' \href{http://dx.doi.org/10.1088/1367-2630/12/8/083053}{{\em New Journal of Physics} {\bfseries 12} (2010) 083053}.

\bibitem{Afanasev:2016}
A.~Afanasev, C.~E. Carlson, and A.~Mukherjee, ``{\it High-multipole excitations of hydrogen-like atoms by twisted photons near a phase singularity},'' \href{http://dx.doi.org/10.1088/2040-8978/18/7/074013}{{\em Journal of Optics} {\bfseries 18} (2016) 074013}, [\href{http://arxiv.org/abs/1602.06640}{{\ttfamily arXiv:1602.06640}} [quant-ph]].

\bibitem{Serbo:2015kia}
V.~Serbo, I.~P. Ivanov, S.~Fritzsche, D.~Seipt, and A.~Surzhykov, ``{\it {Scattering of twisted relativistic electrons by atoms}},'' \href{http://dx.doi.org/10.1103/PhysRevA.92.012705}{{\em Phys. Rev. A} {\bfseries 92} (2015) 012705}, [\href{http://arxiv.org/abs/1505.02587}{{\ttfamily arXiv:1505.02587}} [physics.atom-ph]].

\bibitem{Karlovets:2016uhb}
D.~V. Karlovets, G.~L. Kotkin, V.~G. Serbo, and A.~Surzhykov, ``{\it {Scattering of twisted electron wave packets by atoms in the Born approximation}},'' \href{http://dx.doi.org/10.1103/PhysRevA.95.032703}{{\em Phys. Rev. A} {\bfseries 95} (2017) 032703}, [\href{http://arxiv.org/abs/1612.08252}{{\ttfamily arXiv:1612.08252}} [quant-ph]].

\bibitem{Karlovets:2015nva}
D.~V. Karlovets, G.~L. Kotkin, and V.~G. Serbo, ``{\it {Born approximation for scattering of wave packets on atoms. I. Theoretical background for scattering of a wave packet on a potential field}},'' \href{http://dx.doi.org/10.1103/PhysRevA.92.052703}{{\em Phys. Rev. A} {\bfseries 92} (2015) 052703}, [\href{http://arxiv.org/abs/1508.00026}{{\ttfamily arXiv:1508.00026}} [quant-ph]].

\bibitem{Maiorova:2018inm}
A.~V. Maiorova, S.~Fritzsche, R.~A. Mueller, and A.~Surzhykov, ``{\it {Elastic scattering of twisted electrons by diatomic molecules}},'' \href{http://dx.doi.org/10.1103/PhysRevA.98.042701}{{\em Phys. Rev. A} {\bfseries 98} (2018) 042701}, [\href{http://arxiv.org/abs/1807.06389}{{\ttfamily arXiv:1807.06389}} [physics.atom-ph]].

\bibitem{Mandal:2020ycl}
A.~Mandal, N.~Dhankhar, D.~S\'ebilleau, and R.~Choubisa, ``{\it {Semirelativistic ($e,2e$) study with a twisted electron beam on Cu and Ag}},'' \href{http://dx.doi.org/10.1103/PhysRevA.104.052818}{{\em Phys. Rev. A} {\bfseries 104} (2021) 052818}, [\href{http://arxiv.org/abs/2003.06459}{{\ttfamily arXiv:2003.06459}} [physics.atom-ph]].

\bibitem{Groshev:2019uqa}
M.~E. Groshev, V.~A. Zaytsev, V.~A. Yerokhin, and V.~M. Shabaev, ``{\it {Bremsstrahlung from twisted electrons in the field of heavy nuclei}},'' \href{http://dx.doi.org/10.1103/PhysRevA.101.012708}{{\em Phys. Rev. A} {\bfseries 101} (2020) 012708}, [\href{http://arxiv.org/abs/1911.06162}{{\ttfamily arXiv:1911.06162}} [physics.atom-ph]].

\bibitem{Silenko:2019dfx}
A.~J. Silenko, P.~Zhang, and L.~Zou, ``{\it {Electric Quadrupole Moment and the Tensor Magnetic Polarizability of Twisted Electrons and a Potential for their Measurements}},'' \href{http://dx.doi.org/10.1103/PhysRevLett.122.063201}{{\em Phys. Rev. Lett.} {\bfseries 122} (2019) 063201}, [\href{http://arxiv.org/abs/1902.06882}{{\ttfamily arXiv:1902.06882}} [quant-ph]].

\bibitem{Hayrapetyan:2014faa}
A.~G. Hayrapetyan, O.~Matula, A.~Aiello, A.~Surzhykov, and S.~Fritzsche, ``{\it {Interaction of Relativistic Electron-Vortex Beams with Few-Cycle Laser Pulses}},'' \href{http://dx.doi.org/10.1103/PhysRevLett.112.134801}{{\em Phys. Rev. Lett.} {\bfseries 112} (2014) 134801}.

\bibitem{Bandyopadhyay:2015eri}
P.~Bandyopadhyay, B.~Basu, and D.~Chowdhury, ``{\it {Relativistic Electron Vortex Beams in a Laser Field}},'' \href{http://dx.doi.org/10.1103/PhysRevLett.115.194801}{{\em Phys. Rev. Lett.} {\bfseries 115} (2015) 194801}, [\href{http://arxiv.org/abs/1510.03418}{{\ttfamily arXiv:1510.03418}} [quant-ph]].

\bibitem{Aleksandrov:2022fmp}
I.~A. Aleksandrov, D.~A. Tumakov, A.~Kudlis, V.~A. Zaytsev, and N.~N. Rosanov, ``{\it {Scattering of a twisted electron wavepacket by a finite laser pulse}},'' \href{http://dx.doi.org/10.1103/PhysRevA.106.033119}{{\em Phys. Rev. A} {\bfseries 106} (2022) 033119}, [\href{http://arxiv.org/abs/2206.00110}{{\ttfamily arXiv:2206.00110}} [quant-ph]].

\bibitem{Wu:2021trm}
Y.~Wu, S.~Gargiulo, F.~Carbone, C.~H. Keitel, and A.~P\'alffy, ``{\it {Dynamical Control of Nuclear Isomer Depletion via Electron Vortex Beams}},'' \href{http://dx.doi.org/10.1103/PhysRevLett.128.162501}{{\em Phys. Rev. Lett.} {\bfseries 128} (2022) 162501}, [\href{http://arxiv.org/abs/2107.12448}{{\ttfamily arXiv:2107.12448}} [physics.atom-ph]].

\bibitem{Lu:2023wrf}
Z.-W. Lu {\em et~al.}, ``{\it {Manipulation of Giant Multipole Resonances via Vortex \ensuremath{\gamma} Photons}},'' \href{http://dx.doi.org/10.1103/PhysRevLett.131.202502}{{\em Phys. Rev. Lett.} {\bfseries 131} (2023) 202502}, [\href{http://arxiv.org/abs/2306.08377}{{\ttfamily arXiv:2306.08377}} [nucl-th]].

\bibitem{Sarenac:2024tnj}
D.~Sarenac {\em et~al.}, ``{\it {Small-angle scattering interferometry with neutron orbital angular momentum states}},'' \href{http://dx.doi.org/10.1038/s41467-024-54991-8}{{\em Nature Commun.} {\bfseries 15} no.~1, (2024) 10785}.

\bibitem{Lu:2024gha}
Z.-W. Lu, L.~Guo, M.~Ababekri, J.-l. Zhang, X.-F. Weng, Y.~Wu, Y.-F. Niu, and J.-X. Li, ``{\it {Angular Momentum Resolved Inelastic Electron Scattering for Nuclear Giant Resonances}},'' \href{http://dx.doi.org/10.1103/PhysRevLett.134.052501}{{\em Phys. Rev. Lett.} {\bfseries 134} (2025) 052501}, [\href{http://arxiv.org/abs/2406.05414}{{\ttfamily arXiv:2406.05414}} [nucl-th]].

\bibitem{Ivanov:2019vxe}
I.~P. Ivanov, N.~Korchagin, A.~Pimikov, and P.~Zhang, ``{\it {Doing spin physics with unpolarized particles}},'' \href{http://dx.doi.org/10.1103/PhysRevLett.124.192001}{{\em Phys. Rev. Lett.} {\bfseries 124} (2020) 192001}, [\href{http://arxiv.org/abs/1911.08423}{{\ttfamily arXiv:1911.08423}} [hep-ph]].

\bibitem{Xu:2024jlt}
Y.~Xu, D.~L. Balabanski, V.~Baran, C.~Iorga, and C.~Matei, ``{\it {Vortex photon induced nuclear reaction: Mechanism, model, and application to the studies of giant resonance and astrophysical reaction rate}},'' \href{http://dx.doi.org/10.1016/j.physletb.2024.138622}{{\em Phys. Lett. B} {\bfseries 852} (2024) 138622}.

\bibitem{Afanasev:2021fda}
A.~Afanasev and C.~E. Carlson, ``{\it {Delta Baryon Photoproduction with Twisted Photons}},'' \href{http://dx.doi.org/10.1002/andp.202100228}{{\em Annalen Phys.} {\bfseries 534} no.~3, (2022) 2100228}.

\bibitem{Bu:2021ebc}
Z.~Bu, L.~Ji, S.~Lei, H.~Hu, X.~Zhang, and B.~Shen, ``{\it {Twisted Breit-Wheeler electron-positron pair creation via vortex gamma photons}},'' \href{http://dx.doi.org/10.1103/PhysRevResearch.3.043159}{{\em Phys. Rev. Res.} {\bfseries 3} (2021) 043159}, [\href{http://arxiv.org/abs/2107.06502}{{\ttfamily arXiv:2107.06502}} [hep-ph]].

\bibitem{Lei:2021eqe}
S.~Lei, Z.~Bu, W.~Wang, B.~Shen, and L.~Ji, ``{\it {Generation of relativistic positrons carrying intrinsic orbital angular momentum}},'' \href{http://dx.doi.org/10.1103/PhysRevD.104.076025}{{\em Phys. Rev. D} {\bfseries 104} (2021) 076025}, [\href{http://arxiv.org/abs/2109.02234}{{\ttfamily arXiv:2109.02234}} [hep-ph]].

\bibitem{Sherwin:2017vsf}
J.~A. Sherwin, ``{\it {Theoretical study of the double Compton effect with twisted photons}},'' \href{http://dx.doi.org/10.1103/PhysRevA.95.052101}{{\em Phys. Rev. A} {\bfseries 95} (2017) 052101}.

\bibitem{Maruyama:2017ptl}
T.~Maruyama, T.~Hayakawa, and T.~Kajino, ``{\it {Compton Scattering of $\gamma$-Ray Vortex with Laguerre Gaussian Wave Function}},'' \href{http://dx.doi.org/10.1038/s41598-018-37096-3}{{\em Sci. Rep.} {\bfseries 9} no.~1, (2019) 51}, [\href{http://arxiv.org/abs/1710.09369}{{\ttfamily arXiv:1710.09369}} [hep-ph]].

\bibitem{Bliokh:2017sdz}
K.~Y. Bliokh, M.~R. Dennis, and F.~Nori, ``{\it {Position, spin, and orbital angular momentum of a relativistic electron}},'' \href{http://dx.doi.org/10.1103/PhysRevA.96.023622}{{\em Phys. Rev. A} {\bfseries 96} (2017) 023622}, [\href{http://arxiv.org/abs/1706.01658}{{\ttfamily arXiv:1706.01658}} [quant-ph]].

\bibitem{Silenko:2019okz}
A.~J. Silenko, P.~Zhang, and L.~Zou, ``{\it {Relativistic quantum-mechanical description of twisted paraxial electron and photon beams}},'' \href{http://dx.doi.org/10.1103/PhysRevA.100.030101}{{\em Phys. Rev. A} {\bfseries 100} (2019) 030101}, [\href{http://arxiv.org/abs/1904.02434}{{\ttfamily arXiv:1904.02434}} [quant-ph]].

\bibitem{Silenko:2018eed}
A.~J. Silenko, P.~Zhang, and L.~Zou, ``{\it {Relativistic quantum dynamics of twisted electron beams in arbitrary electric and magnetic fields}},'' \href{http://dx.doi.org/10.1103/PhysRevLett.121.043202}{{\em Phys. Rev. Lett.} {\bfseries 121} (2018) 043202}, [\href{http://arxiv.org/abs/1803.07410}{{\ttfamily arXiv:1803.07410}} [physics.class-ph]].

\bibitem{vanKruining:2017anw}
K.~van Kruining, A.~G. Hayrapetyan, and J.~B. G\"otte, ``{\it {Nonuniform Currents and Spins of Relativistic Electron Vortices in a Magnetic Field}},'' \href{http://dx.doi.org/10.1103/PhysRevLett.119.030401}{{\em Phys. Rev. Lett.} {\bfseries 119} (2017) 030401}, [\href{http://arxiv.org/abs/1702.05271}{{\ttfamily arXiv:1702.05271}} [quant-ph]].

\bibitem{Barnett:2017wrr}
S.~M. Barnett, ``{\it {Relativistic Electron Vortices}},'' \href{http://dx.doi.org/10.1103/PhysRevLett.118.114802}{{\em Phys. Rev. Lett.} {\bfseries 118} (2017) 114802}.

\bibitem{Bialynicki-Birula:2016unl}
I.~Bialynicki-Birula and Z.~Bialynicka-Birula, ``{\it {Relativistic Electron Wave Packets Carrying Angular Momentum}},'' \href{http://dx.doi.org/10.1103/PhysRevLett.118.114801}{{\em Phys. Rev. Lett.} {\bfseries 118} (2017) 114801}, [\href{http://arxiv.org/abs/1611.04445}{{\ttfamily arXiv:1611.04445}} [quant-ph]].

\bibitem{Bu:2024}
Z.~Bu, L.~Ji, X.~Geng, S.~Liu, S.~Lei, B.~Shen, R.~Li, and Z.~Xu, ``{\it Generation of quantum vortex electrons with intense laser pulses},'' \href{http://dx.doi.org/10.1002/advs.202404564}{{\em Advanced Science} {\bfseries 11} no.~41, (2024) 2404564}.

\bibitem{Li:2024mzd}
Z.~Li, S.~Liu, B.~Liu, L.~Ji, and I.~P. Ivanov, ``{\it {Unambiguous Detection of High-Energy Vortex States via the Superkick Effect}},'' \href{http://dx.doi.org/10.1103/PhysRevLett.133.265001}{{\em Phys. Rev. Lett.} {\bfseries 133} (2024) 265001}, [\href{http://arxiv.org/abs/2406.06795}{{\ttfamily arXiv:2406.06795}} [hep-ph]].

\bibitem{Katoh:2016yqc}
M.~Katoh {\em et~al.}, ``{\it {Twisted Radiation by Electrons in Spiral Motion}},'' \href{http://dx.doi.org/10.1038/s41598-017-06442-2}{{\em Sci. Rep.} {\bfseries 7} (2017) 6130}, [\href{http://arxiv.org/abs/1609.03869}{{\ttfamily arXiv:1609.03869}} [physics.optics]].

\bibitem{Petrillo:2016}
V.~Petrillo, G.~Dattoli, I.~Drebot, and F.~Nguyen, ``{\it Compton scattered x-gamma rays with orbital momentum},'' \href{http://dx.doi.org/10.1103/PhysRevLett.117.123903}{{\em Phys. Rev. Lett.} {\bfseries 117} (2016) 123903}.

\bibitem{Jentschura:2010ap}
U.~D. Jentschura and V.~G. Serbo, ``{\it {Generation of High-Energy Photons with Large Orbital Angular Momentum by Compton Backscattering}},'' \href{http://dx.doi.org/10.1103/PhysRevLett.106.013001}{{\em Phys. Rev. Lett.} {\bfseries 106} (2011) 013001}, [\href{http://arxiv.org/abs/1008.4788}{{\ttfamily arXiv:1008.4788}} [physics.acc-ph]].

\bibitem{Jentschura:2011ih}
U.~D. Jentschura and V.~G. Serbo, ``{\it {Compton Upconversion of Twisted Photons: Backscattering of Particles with Non-Planar Wave Functions}},'' \href{http://dx.doi.org/10.1140/epjc/s10052-011-1571-z}{{\em Eur. Phys. J. C} {\bfseries 71} (2011) 1571}, [\href{http://arxiv.org/abs/1101.1206}{{\ttfamily arXiv:1101.1206}} [physics.acc-ph]].

\bibitem{Katoh:2016aww}
M.~Katoh, M.~Fujimoto, H.~Kawaguchi, K.~Tsuchiya, K.~Ohmi, T.~Kaneyasu, Y.~Taira, M.~Hosaka, A.~Mochihashi, and Y.~Takashima, ``{\it {Angular Momentum of Twisted Radiation from an Electron in Spiral Motion}},'' \href{http://dx.doi.org/10.1103/PhysRevLett.118.094801}{{\em Phys. Rev. Lett.} {\bfseries 118} (2017) 094801}, [\href{http://arxiv.org/abs/1610.02182}{{\ttfamily arXiv:1610.02182}} [physics.optics]].

\bibitem{Taira:2017}
Y.~{Taira}, T.~{Hayakawa}, and M.~{Katoh}, ``{\it {Gamma-ray vortices from nonlinear inverse Thomson scattering of circularly polarized light}},'' \href{http://dx.doi.org/10.1038/s41598-017-05187-2}{{\em Scientific Reports} {\bfseries 7} (2017) 5018}.

\bibitem{Chen:2018tkb}
Y.-Y. Chen, J.-X. Li, K.~Z. Hatsagortsyan, and C.~H. Keitel, ``{\it {\ensuremath{\gamma} -Ray Beams with Large Orbital Angular Momentum via Nonlinear Compton Scattering with Radiation Reaction}},'' \href{http://dx.doi.org/10.1103/PhysRevLett.121.074801}{{\em Phys. Rev. Lett.} {\bfseries 121} (2018) 074801}, [\href{http://arxiv.org/abs/1802.04748}{{\ttfamily arXiv:1802.04748}} [physics.plasm-ph]].

\bibitem{Ababekri:2022mob}
M.~Ababekri, R.-T. Guo, F.~Wan, B.~Qiao, Z.~Li, C.~Lv, B.~Zhang, W.~Zhou, Y.~Gu, and J.-X. Li, ``{\it {Vortex \ensuremath{\gamma} photon generation via spin-to-orbital angular momentum transfer in nonlinear Compton scattering}},'' \href{http://dx.doi.org/10.1103/PhysRevD.109.016005}{{\em Phys. Rev. D} {\bfseries 109} (2024) 016005}, [\href{http://arxiv.org/abs/2211.05467}{{\ttfamily arXiv:2211.05467}} [hep-ph]].

\bibitem{Guo:2023uyu}
R.-T. Guo, M.~Ababekri, Q.~Zhao, Y.~I. Salamin, L.-L. Ji, Z.-G. Bu, Z.-F. Xu, X.-F. Weng, and J.-X. Li, ``{\it {Generation of \ensuremath{\gamma} photons with extremely large orbital angular momenta}},'' \href{http://dx.doi.org/10.1103/PhysRevA.110.032202}{{\em Phys. Rev. A} {\bfseries 110} (2024) 032202}, [\href{http://arxiv.org/abs/2310.16306}{{\ttfamily arXiv:2310.16306}} [hep-ph]].

\bibitem{Bogdanov:2019ocq}
O.~V. Bogdanov, P.~O. Kazinski, and G.~Y. Lazarenko, ``{\it {Semiclassical probability of radiation of twisted photons in the ultrarelativistic limit}},'' \href{http://dx.doi.org/10.1103/PhysRevD.99.116016}{{\em Phys. Rev. D} {\bfseries 99} (2019) 116016}, [\href{http://arxiv.org/abs/1903.04024}{{\ttfamily arXiv:1903.04024}} [physics.acc-ph]].

\bibitem{Jiang:2024fit}
J.-J. Jiang, K.-H. Zhuang, J.-D. Chen, J.-X. Li, and Y.-Y. Chen, ``{\it {Controlling the polarization and vortex charge of $\gamma$ photons via nonlinear Compton scattering}},'' [\href{http://arxiv.org/abs/2410.20658}{{\ttfamily arXiv:2410.20658}} [hep-ph]].

\bibitem{Wang:2020}
J.~Wang, X.~Li, L.~Gan, Y.~Xie, C.~Zhong, C.~Zhou, S.~Zhu, X.~He, and B.~Qiao, ``{\it Generation of intense vortex gamma rays via spin-to-orbital conversion of angular momentum in relativistic laser-plasma interactions},'' \href{http://dx.doi.org/10.1103/PhysRevApplied.14.014094}{{\em Phys. Rev. Appl.} {\bfseries 14} (2020) 014094}.

\bibitem{liu:2016}
C.~Liu, B.~Shen, X.~Zhang, Y.~Shi, L.~Ji, W.~Wang, L.~Yi, L.~Zhang, T.~Xu, Z.~Pei, {\em et~al.}, ``{\it Generation of gamma-ray beam with orbital angular momentum in the qed regime},'' \href{http://dx.doi.org/10.1063/1.4963396}{{\em Physics of Plasmas} {\bfseries 23} (2016) 093120}.

\bibitem{hu:2021}
Y.-T. Hu, J.~Zhao, H.~Zhang, Y.~Lu, W.-Q. Wang, L.-X. Hu, F.-Q. Shao, and T.-P. Yu, ``{\it Attosecond $\gamma$-ray vortex generation in near-critical-density plasma driven by twisted laser pulses},'' \href{http://dx.doi.org/10.1063/5.0028203}{{\em Applied Physics Letters} {\bfseries 118} (2021) 054101}.

\bibitem{ju:2018}
L.~Ju, C.~Zhou, K.~Jiang, T.~Huang, H.~Zhang, T.~Cai, J.~Cao, B.~Qiao, and S.~Ruan, ``{\it Manipulating the topological structure of ultrarelativistic electron beams using laguerre--gaussian laser pulse},'' \href{http://dx.doi.org/10.1088/1367-2630/aac68a}{{\em New Journal of Physics} {\bfseries 20} (2018) 063004}.

\bibitem{Karlovets:2020tlg}
D.~Karlovets, ``{\it {Vortex particles in axially symmetric fields and applications of the quantum Busch theorem}},'' \href{http://dx.doi.org/10.1088/1367-2630/abeacc}{{\em New J. Phys.} {\bfseries 23} (2021) 033048}, [\href{http://arxiv.org/abs/2012.05741}{{\ttfamily arXiv:2012.05741}} [quant-ph]].

\bibitem{Karlovets:2022evc}
D.~V. Karlovets, S.~S. Baturin, G.~Geloni, G.~K. Sizykh, and V.~G. Serbo, ``{\it {Generation of vortex particles via generalized measurements}},'' \href{http://dx.doi.org/10.1140/epjc/s10052-022-10991-w}{{\em Eur. Phys. J. C} {\bfseries 82} no.~11, (2022) 1008}, [\href{http://arxiv.org/abs/2201.07997}{{\ttfamily arXiv:2201.07997}} [hep-ph]].

\bibitem{Karlovets:2022mhb}
D.~V. Karlovets, S.~S. Baturin, G.~Geloni, G.~K. Sizykh, and V.~G. Serbo, ``{\it {Shifting physics of vortex particles to higher energies via quantum entanglement}},'' \href{http://dx.doi.org/10.1140/epjc/s10052-023-11529-4}{{\em Eur. Phys. J. C} {\bfseries 83} no.~5, (2023) 372}, [\href{http://arxiv.org/abs/2203.12012}{{\ttfamily arXiv:2203.12012}} [hep-ph]].

\bibitem{Ivanov:2011kk}
I.~P. Ivanov, ``{\it {Colliding particles carrying non-zero orbital angular momentum}},'' \href{http://dx.doi.org/10.1103/PhysRevD.83.093001}{{\em Phys. Rev. D} {\bfseries 83} (2011) 093001}, [\href{http://arxiv.org/abs/1101.5575}{{\ttfamily arXiv:1101.5575}} [hep-ph]].

\bibitem{Laudau:1982}
V.~B. Berestetskii, E.~M. Lifshitz, and L.~P. Pitaevskii, {\em Quantum Electrodynamics (Second Edition)}.
\newblock Butterworth-Heinemann, Oxford, 1982.

\bibitem{Brodsky:1997de}
S.~J. Brodsky, H.-C. Pauli, and S.~S. Pinsky, ``{\it {Quantum chromodynamics and other field theories on the light cone}},'' \href{http://dx.doi.org/10.1016/S0370-1573(97)00089-6}{{\em Phys. Rept.} {\bfseries 301} (1998) 299--486}, [\href{http://arxiv.org/abs/hep-ph/9705477}{{\ttfamily arXiv:hep-ph/9705477}}].

\bibitem{Seipt:2020diz}
D.~Seipt and B.~King, ``{\it {Spin- and polarization-dependent locally-constant-field-approximation rates for nonlinear Compton and Breit-Wheeler processes}},'' \href{http://dx.doi.org/10.1103/PhysRevA.102.052805}{{\em Phys. Rev. A} {\bfseries 102} (2020) 052805}, [\href{http://arxiv.org/abs/2007.11837}{{\ttfamily arXiv:2007.11837}} [physics.plasm-ph]].

\bibitem{King:2020btz}
B.~King and S.~Tang, ``{\it {Nonlinear Compton scattering of polarized photons in plane-wave backgrounds}},'' \href{http://dx.doi.org/10.1103/PhysRevA.102.022809}{{\em Phys. Rev. A} {\bfseries 102} (2020) 022809}, [\href{http://arxiv.org/abs/2003.01749}{{\ttfamily arXiv:2003.01749}} [hep-ph]].

\bibitem{Karlovets:2012eu}
D.~V. Karlovets, ``{\it {Electron with orbital angular momentum in a strong laser wave}},'' \href{http://dx.doi.org/10.1103/PhysRevA.86.062102}{{\em Phys. Rev. A} {\bfseries 86} (2012) 062102}, [\href{http://arxiv.org/abs/1206.6622}{{\ttfamily arXiv:1206.6622}} [hep-ph]].

\bibitem{Ivanov:2011tm}
I.~P. Ivanov, ``{\it {Non-forward scattering of twisted particles}},'' [\href{http://arxiv.org/abs/1101.1630}{{\ttfamily arXiv:1101.1630}} [hep-ph]].

\bibitem{Ivanov:2011bv}
I.~P. Ivanov and V.~G. Serbo, ``{\it {Scattering of Twisted Particles: Extension to Wave Packets and orbital helicity}},'' \href{http://dx.doi.org/10.1103/PhysRevA.84.033804}{{\em Phys. Rev. A} {\bfseries 84} (2011) 033804}, [\href{http://arxiv.org/abs/1105.6244}{{\ttfamily arXiv:1105.6244}} [hep-ph]].

\bibitem{Ponce_de_Leon:2015}
J.~{Ponce de Leon}, ``{\it {Revisiting the orthogonality of Bessel functions of the first kind on an infinite interval}},'' \href{http://dx.doi.org/10.1088/0143-0807/36/1/015016}{{\em European Journal of Physics} {\bfseries 36} (2015) 015016}.

\bibitem{Gradshteyn:2007}
I.~Gradshteyn and I.~Ryzhik, {\em Table of integrals, series, and products}.
\newblock Academic Press, 2007.

\bibitem{Kotkin:1992bj}
G.~L. Kotkin, V.~G. Serbo, and A.~Schiller, ``{\it {Processes with large impact parameters at colliding beams}},'' \href{http://dx.doi.org/10.1142/S0217751X92002131}{{\em Int. J. Mod. Phys. A} {\bfseries 7} (1992) 4707--4745}.

\bibitem{Gonoskov:2021hwf}
A.~Gonoskov, T.~G. Blackburn, M.~Marklund, and S.~S. Bulanov, ``{\it {Charged particle motion and radiation in strong electromagnetic fields}},'' \href{http://dx.doi.org/10.1103/RevModPhys.94.045001}{{\em Rev. Mod. Phys.} {\bfseries 94} (2022) 045001}, [\href{http://arxiv.org/abs/2107.02161}{{\ttfamily arXiv:2107.02161}} [physics.plasm-ph]].

\bibitem{Bliokh:2011fi}
K.~Y. Bliokh, M.~R. Dennis, and F.~Nori, ``{\it {Relativistic Electron Vortex Beams: Angular Momentum and Spin-Orbit Interaction}},'' \href{http://dx.doi.org/10.1103/PhysRevLett.107.174802}{{\em Phys. Rev. Lett.} {\bfseries 107} (2011) 174802}, [\href{http://arxiv.org/abs/1105.0331}{{\ttfamily arXiv:1105.0331}} [quant-ph]].

\end{thebibliography}\endgroup
\bibliographystyle{paper250406style}

\end{document}